\pdfoutput=1
\documentclass[structabstract]{aa}
\usepackage{txfonts}
\usepackage{natbib}
\usepackage{epsfig}
\usepackage{graphicx}
\usepackage{natbib}
\bibpunct{(}{)}{;}{a}{}{,} % to follow the A&A style
\usepackage{}
\usepackage{subfigure}
\usepackage{latexsym} 
\usepackage{pifont}
\input epsf
\newcount\Comments  % 0 suppresses notes to selves in text
\usepackage[usenames,dvipsnames]{xcolor}
\newcommand{\kibitz}[2]{\ifnum\Comments=1\textcolor{#1}{#2}\fi}

\newcommand{\xmm}{{\it XMM~\/}}
\newcommand{\xmmn}{{\it XMM-Newton~\/}}

\newcommand{\chandra}{{\it Chandra~\/}}

\newcommand{\pwn}{{pulsar wind nebula~\/}}
\newcommand{\obs}{{observation~\/}}
\newcommand{\obss}{{observations~\/}}

\newcommand{\psr}{{PSR\,J0855$-$4644}}
\def\ergcms{{\rm ~erg~cm^{-2}~s^{-1}}}

\def\ergsec{{\rm ~erg~s^{-1}}}
\def\cm2{{\rm ~cm^{-2}}}

\def\dg{^{\circ}}
\def\H0{{\rm ~km~s^{-1}~Mpc^{-1}}}

\def\ie{{\it i.e.~\/}}

\def\la{\mathrel{\hbox{\rlap{\hbox{\lower4pt\hbox{$\sim$}}}{\raise2pt\hbox{$<$}}}}}
\def\ga{\mathrel{\hbox{\rlap{\hbox{\lower4pt\hbox{$\sim$}}}{\raise2pt\hbox{$>$}}}}}
\def\d25{D$_{25}$}

\def\deg{\hbox{$^\circ$}}
\def\arcm{\hbox{$^\prime$~\/}}
\def\arcs{\hbox{$^{\prime\prime}$}}

\def\eps@scaling{1.0}%
\newcommand\plottwo[2]{%\input{psrj0855_draft_v4.tex}

  \centering
  \leavevmode
  \columnwidth=.45\columnwidth
  \includegraphics[width={\eps@scaling\columnwidth}]{#1}%
  \hfil
  \includegraphics[width={\eps@scaling\columnwidth}]{#2}%
}%

\begin{document}

\title{Constraining the geometry of \psr: %Confirmation of 
A nearby Pulsar Wind Nebula with Double Torus/Jet Morphology}
\titlerunning{The compact PWN around PSR J0855-4644}
\author{C.~Maitra\inst{1}\thanks{Present address: Max-Planck-Institut f{\"ur} extraterrestrische Physik, Giessenbachstra{\ss}e, 85748 Garching, Germany}
\and F.~Acero\inst{1}
\and C.~Venter\inst{2}
}

\institute{Laboratoire AIM, IRFU/Service d'Astrophysique - CEA/DRF - CNRS - Universite Paris Diderot, Bat. 709, CEA-Saclay, 91191 Gif-sur-Yvette Cedex, France 
\email{cmaitra@mpg.mpe.de} 
\and 
Centre for Space Research, North-West University, Potchefstroom Campus, Private Bag X6001, Potchefstroom 2520, South Africa
} 

\date{Received ... / Accepted ...}

\abstract 
{} 
{\psr~is a fast-spinning, energetic pulsar discovered at radio wavelengths near the south-eastern rim of the supernova remnant RX\,J0852.0$-$4622. A follow-up \xmmn observation revealed the pulsar's X-ray counterpart and a slightly  asymmetric \pwn suggesting possible jet structures. Lying at a distance $d\leq 900$~pc, PSR\,J0855$-$4644 is a pulsar with one of the highest $\dot{E}$/$d^{2}$ from which no GeV $\gamma$-ray pulsations have been detected. With a dedicated \chandra \obs we aim to further resolve the possible jet structures of the nebula and study the pulsar's geometry in order to understand the lack of $\gamma$-ray pulsations.}
{We perform detailed spatial modelling to constrain the geometry of the \pwn and in particular the pulsar's line of sight (observer angle) $\zeta_{\rm PSR}$ defined as the angle between the direction of the observer and the pulsar spin axis. We also perform geometric radio and $\gamma$-ray light curve modelling using a hollow-cone radio beam model together with two-pole caustic and outer gap models to further constrain $\zeta_{\rm PSR}$ and the magnetic obliquity $\alpha$ defined as the angle between the magnetic and spin axes of the pulsar.}
{The \chandra \obs reveals that the compact \xmm source, thought to be the X-ray pulsar, can be further resolved into a point source surrounded by an elongated axisymmetric nebula with a longitudinal extent of 10\arcs. The pulsar flux represents only $\sim$ 1\% of the \xmm compact source and its spectrum is well  described by a blackbody of temperature $kT=0.2$~keV while the surrounding nebula has a much harder spectrum ($\Gamma=1.1$ for a power-law model). Assuming the origin of the extended emission is from a double torus yields $\zeta_{\rm PSR}=32.5\dg\pm 4.3\dg$. The detection of thermal X-rays from the pulsar may point to a low value of $\lvert \zeta-\alpha\rvert\ $ if this emission originates from a heated polar cap. Independent constraints from geometric light curve modelling yield $\alpha\lesssim55\dg$ and $\zeta\lesssim55\dg$, and $10\dg\lesssim|\zeta-\alpha|\lesssim30\dg$. A $\chi^2$ fit to the radio light curve yields a best fit at $(\alpha,\zeta_{\rm PSR}) = (22\dg, 8\dg)$, with an alternative fit at $(\alpha,\zeta_{\rm PSR}) = (9\dg, 25\dg)$ within $3\sigma$. The lack of non-thermal X-ray emission from the pulsar further supports low values for $\alpha$ and $\zeta$ under the assumption that X-rays and $\gamma$-rays are generated in the same region of the pulsar magnetosphere. Such a geometry would explain, in the standard caustic pulsar model picture, the radio-loud and $\gamma$-ray-quiet behaviour of this high $\dot{E}$/$d^{2}$ pulsar.}
{}


\maketitle

\section{Introduction}
\label{sec:intro}
Pulsar wind nebulae (PWNe) are powered by young energetic pulsars and are excellent sites to study the energetics of and particle spectrum injected by young rotation-powered pulsars. The unprecedented spatial resolution of \chandra has unravelled highly structured compact PWNe around many of these sources, including equatorial and polar outflows \citep{weisskopf2000,gaensler2006,pwnchandra}. An accurate knowledge of the PWN morphology is crucial to constrain the geometry of the pulsar, especially the pulsar's line of sight $\zeta_{\rm PSR}$ defined as the angle between the direction of the observer and the pulsar spin axis, and the magnetic obliquity $\alpha$ defined as the angle between the magnetic axis and the spin axis of the pulsar. Such constraints can in turn provide useful insights for models of high-energy emission from pulsar magnetospheres, such as the polar-cap (PC) model \citep{1996ApJ...458..278D}, the outer gap (OG) model \citep{CHR86a,1995ApJ...438..314R,Romani96}, and the slot gap or two-pole caustic (TPC) model \citep{Arons83,Dyks03}.

\psr~was discovered by the Parkes multibeam radio survey lying on the south-eastern rim of the RX~J0852.0$-$4622 supernova remnant \citep[SNR; ][]{2003MNRAS.342.1299K}. The measured spin characteristics, such as the spin period $P=65$~ms and its first derivative $\dot{P} =  7.3  \times 10^{-15} ~s ~s^{-1}$ gives a spin-down luminosity $\dot{E} = 1.1 \times 10^{36}$ erg/s (assuming a moment of inertia $I = 10^{45} g ~cm^{2}$  for standard neutron star parameters), and characteristic age $\tau_{\rm c} \equiv {P}/(2\dot{P}) = 140$ kyrs. A dedicated \xmmn \obs revealed that the X-ray counterpart of the pulsar is surrounded by extended non-thermal emission, the associated PWN \citep{facero2013}. The emission was 150\arcs\ in extent, including two large-scale structures with an angular separation of $\sim$180$\dg$, resembling possible jets. In addition,  comparison of column densities provided an upper limit to the distance of the pulsar \psr~of $d\leq 900$~pc. With this revised distance, \psr~is the second most energetic pulsar, after the Vela pulsar, within a distance of 1 kpc from Earth, and could therefore contribute to the local cosmic-ray e$^{-}$/e$^{+}$ spectrum \citep{facero2013}. This furthermore makes this pulsar the highest $\dot{E}$/$d^{2}$ system not detected  at $\gamma$-ray energies with the \textit{Fermi}-LAT instrument \citep{abdo13-2PC}. As mentioned earlier, determination of the pulsar/PWN geometry holds the key to understand this paradox. 

We present results of a 38 ks on-axis \chandra observation to study the small-scale morphology of the nebula surrounding PSR J0855-4644 and its geometry. With the arcsecond spatial resolution of \chandra we have resolved the compact \xmmn source that was assumed to be the X-ray counterpart of the pulsar into a $\sim$ 10\arcs\ extended emission with jets and a possible double torus. We performed a morphological and spectral analysis, and constrained the observer angle of the system. From geometric light curve modelling, we could also independently constrain the ($\alpha$, $\zeta_{\rm PSR}$) space by comparing our predictions to the radio pulse profile of the pulsar (and taking the non-detection of pulsed $\gamma$ rays into account).  

The \obss and analysis are described in Section 2. Section 3 presents the spatial analysis including the imaging and morphological fitting of the source. Section~4 presents the detailed spectral analysis of the pulsar, the surrounding nebula, and its decomposed regions. Section~5 details our geometric light curve modelling and comparison with observations. Section~6 is the discussion, and Section~7 the summary and future work.
%%%%%%%%%%%%%%

\begin{figure*}
\hspace*{-0.95cm}
\subfigure[]{\includegraphics[angle=0,scale=0.58]{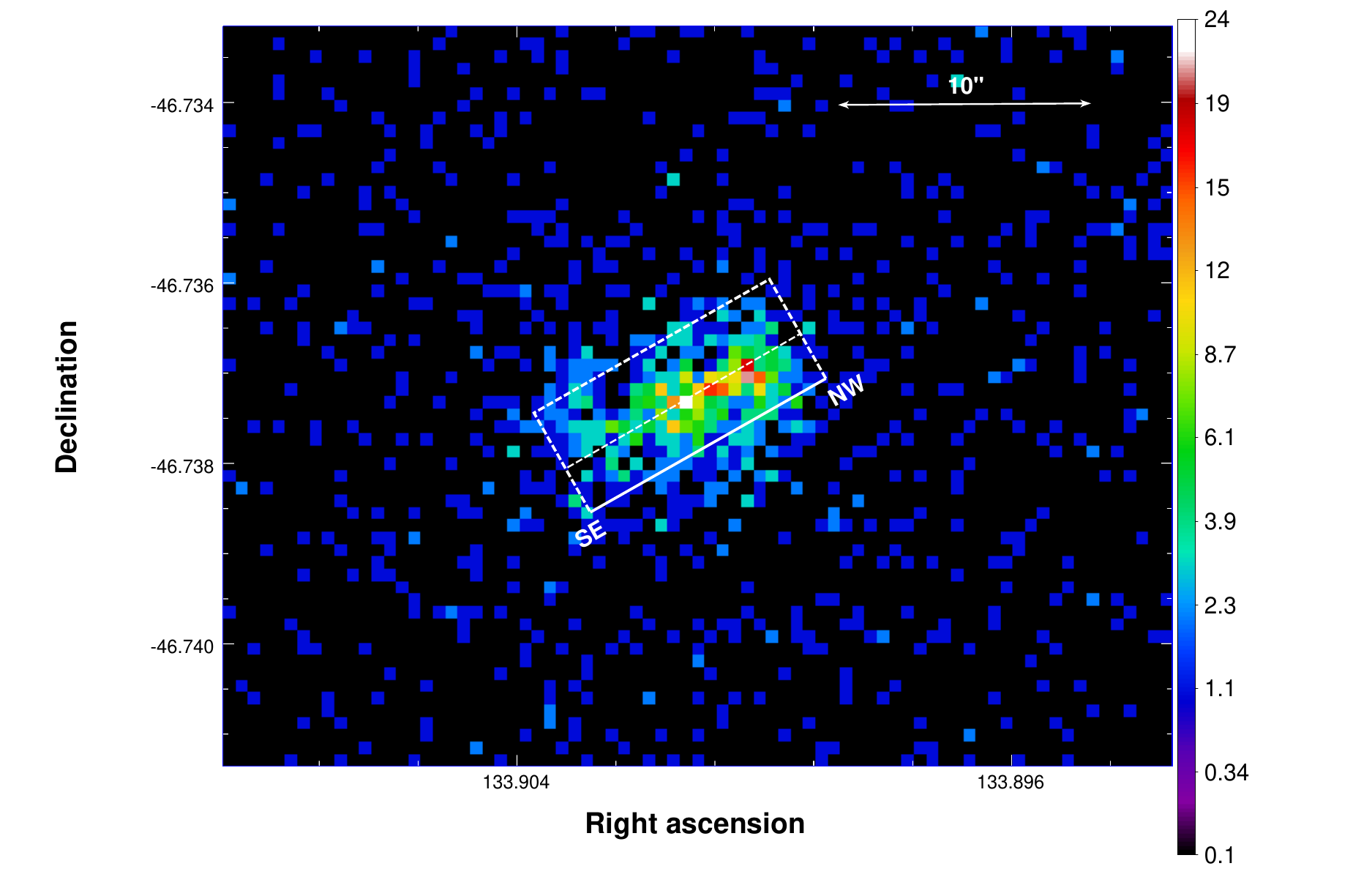}}
\hspace*{-1.5cm}
\subfigure[]{\includegraphics[angle=0,scale=0.58]{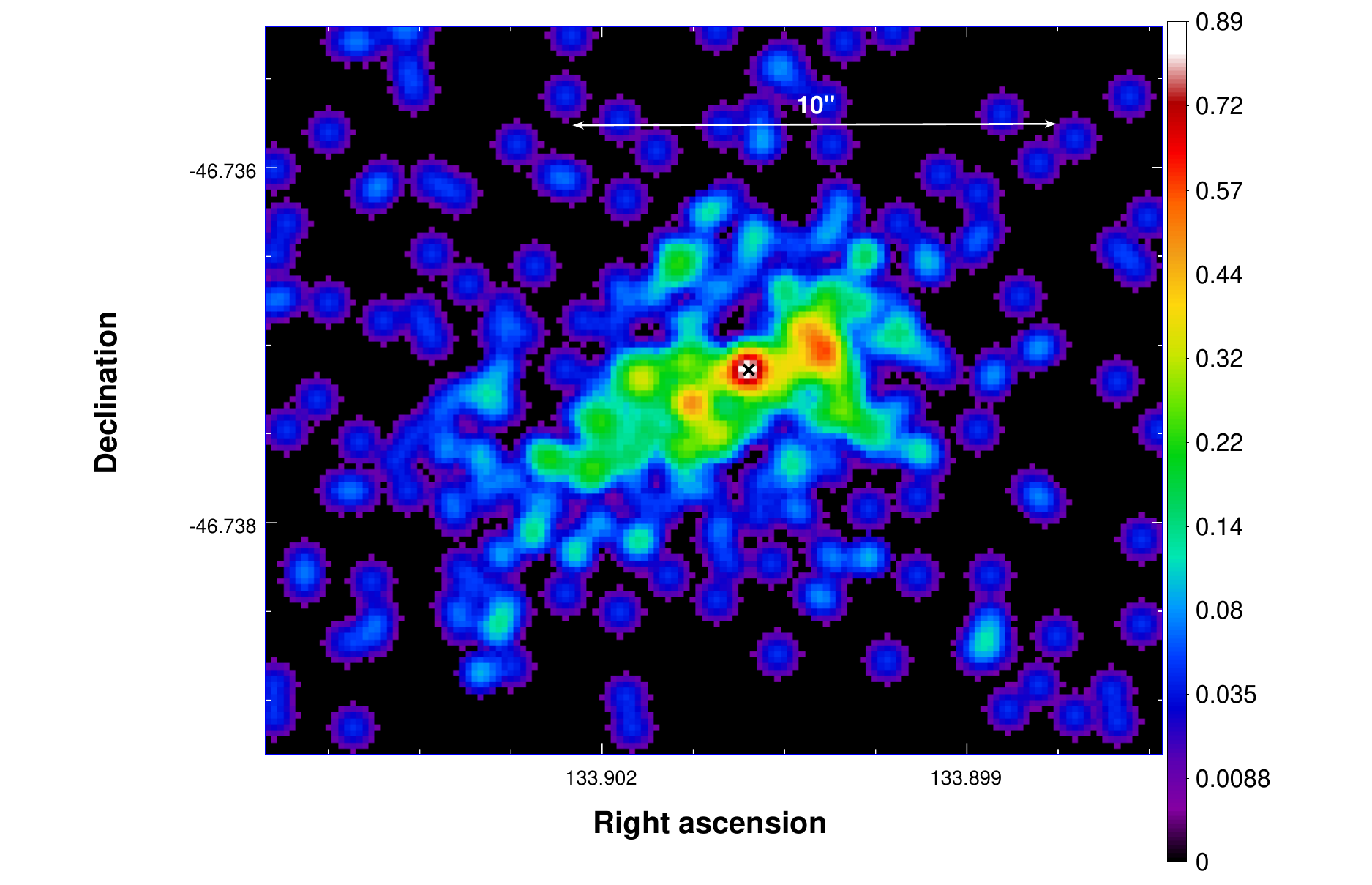}}
\hspace*{-1.6cm}
\subfigure[]{\includegraphics[angle=0,scale=0.58]{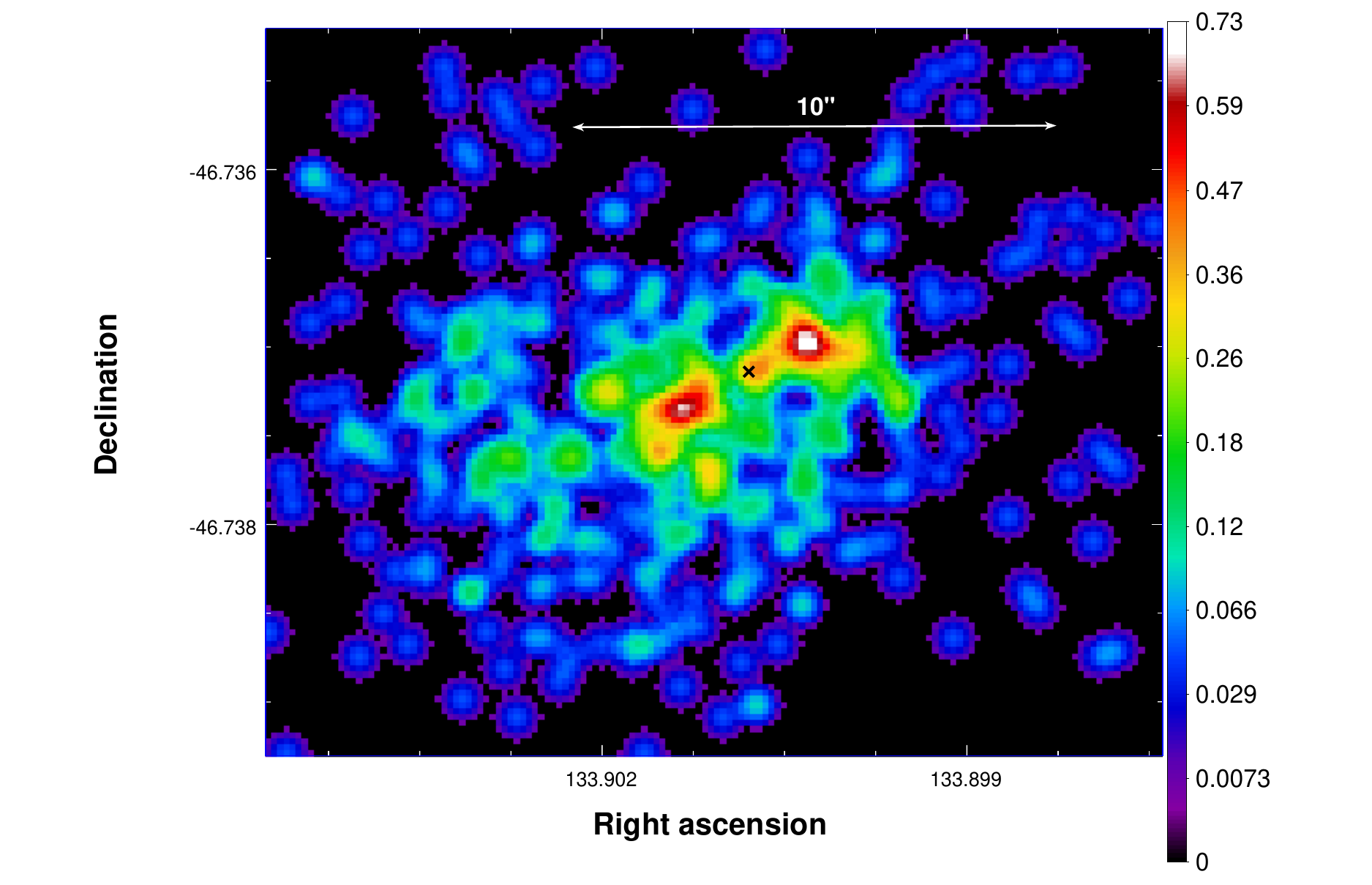}}
%\vspace*{2.0cm}
%\hspace*{0.18 cm}
\subfigure[]{\includegraphics[angle=0,scale=0.53]{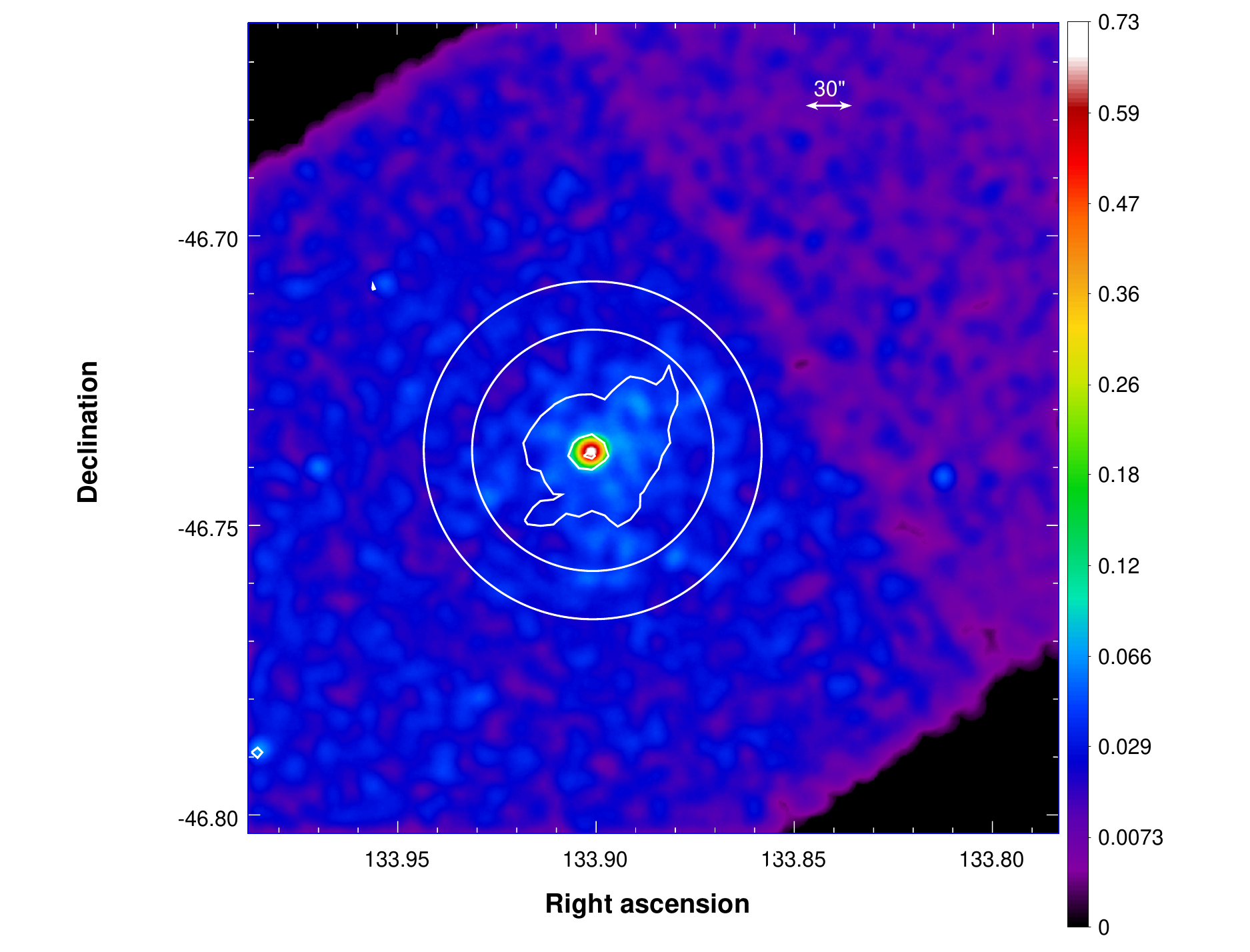}}
\caption{Panel~(a): \chandra ACIS-S (0.5--8 keV) full resolution image of the PWN around PSR J0855$-$4644. The image size is 29\arcs\ $\times$ 27\arcs. The box shows the extraction region for the counts profile in Figure~2. The white dashed line inside shows the symmetry axis along which the nebula is extended. Panel~(b): ACIS-S (0.5--2 keV) sub-pixel image (image size is 18\arcs\ x 15\arcs) produced at $\frac{1}{4}$ resolution. The black cross denotes the radio position of the pulsar. The image has been smoothed with a 1\arcs\ FWHM Gaussian to emphasize the nebular morphology. Panel~(c): Same as in Panel (b), but in the energy range of 2--8 keV. Panel~(d): Larger 2--8 keV image centred on \psr~(ACIS-S3 CCD) smoothed with a Gaussian of 10\arcs. Overlayed are contours from the \xmm~image \citep[Fig. 2 of][]{facero2013} at levels values of 1.13, 6.0, and 14.1 counts arcsecs$^{-2}$. The annular region used for the background spectral extraction (inner radius 75\arcs) is shown by white solid lines. In all the images, the colour bar indicates the square root of number of counts, the $X$ and $Y$ axes in degrees, and units are counts for all the images.}
\label{images}
\end{figure*}
%%%%%%%%%%%%%%%%

\section{Observations and analysis}
\label{sec:obs}
The Advanced CCD Imaging Spectrometer (ACIS) onboard \chandra offers simultaneous high resolution images and moderate-resolution spectra. ACIS consists of 10 planar CCDs, 4 front-illuminated ones (ACIS-I) are arranged in $2 \times 2$ array, and 6 back-illuminated ones (ACIS-S) in a $1 \times 6$ array.
The \chandra \obs (ObsID 13780) was carried out with the ACIS-S as the primary instrument, and was performed in full-frame timed exposure (TE) mode which uses the standard 3.2 s frame time and the FAINT telementry format.  The \obs was performed on 3 September 2012 with \psr ~positioned at the aimpoint of the S3 CCD with an exposure of 38.02 ks. Data reduction was performed using CIAO 4.7, CALDB 4.6.7, and the standard prescribed analysis procedure\footnote{http://cxc.harvard.edu/ciao/}. The event file was reprocessed with \textit{chandra$\_$repro}, which incorporates the subpixel repositioning algorithm EDSER as a default to obtain a better angular resolution for sources near the centre of the field of view. The effective exposure of the observation after filtering was 37.98 ks. We also checked for the possible presence of pileup in the data. Many times, multiple photons impinge on a single pixel during readout, leading to it being read out as one single photon with a large pulse height, and causing photon pileup. Using the tool {\it pileup$\_$map}, the estimated fraction of pileup in the centremost pixel was $<5\%$ and so the effect was not important for our observation. Images were created using the task \textit{dmcopy}. Spatial analysis was performed using the \textit{Sherpa} analysis package 4.4\footnote{http://cxc.harvard.edu/sherpa4.4/}. The task {\it specextract} was used for extracting source and background spectra and response files from regions of interest.
The selection of regions used for spectral extraction is described in Section~4. Spectra were fitted using \textit{XSPEC v12.8.2}\footnote{http://heasarc.gsfc.nasa.gov/xanadu/xspec/}.
 
%%%%%%%%%
Figure~\ref{images}a shows a full resolution \chandra image in the energy range 0.5--8 keV centred on \psr. The point source seen with \xmm that was previously thought to be the pulsar X-ray counterpart is in fact composed of an extended structure elongated in the south-east/north-west direction along a symmetry axis (See Figure~\ref{images}a). This component of the PWN is more compact  ($\sim$ 10\arcs) compared to the much larger diffuse nebula seen with \xmm measuring 150\arcs\ \citep{facero2013}. Signatures of the diffuse nebula can be seen in the \chandra image in the high-energy 2--8 keV band (free of emission from the Vela Senior SNR) after smoothing it at the scale of the compact PWN, \ie 10\arcs\ (See Figure~\ref{images}d). The fine structures of the compact PWN can be seen more clearly from its subpixel images of the same region after dividing them into soft {\it i.e.} 0.5--2 keV and hard, {\it i.e.} 2--8 keV energy bands (See Figures~\ref{images}b and c). The elongated PWN structure is composed of two lobes lying on the symmetry axis, more or less equidistant from the  pulsar in the middle. The pulsar spectrum is soft and almost disappears in the 2--8 keV energy band. The nebular structure on the other hand is much harder and extends more to the east in the form of a tail-like elongation. The structure is reminiscent of axisymmetric features commonly seen in young PWN systems like the equatorial outflows (tori), and collimated polar outflows (jets) \citep{gaensler2006,pwnchandra}. The two lobes symmetric about the pulsar can be explained by the three possible scenarios: a) double-torus b) double-torus and jets, and c) jets-only morphology.  In the first scenario the lobes could be the two rings of a double torus structure as it would appear in sky projection, spaced more or less equidistant from the pulsar centered along the torus axis (which is coincident with the symmetry axis, a.k.a.\ the spin axis).  The brightened central parts of the lobes are due to the doppler boosting of the torus. In the second scenario, the morphology of the PWN around \psr~ is that of a double torus-jet structure, with the presence of both the equatorial torus and polar jet features contributing to the lobes. This is similar to the PWNe systems seen in Vela and PSR~J2021+3651 \citep{helfand2001,hessels2004}. In the third case, the lobes could be the jets along the pulsar spin axis (the symmetry axis in Figure~\ref{images}a). In all three cases, however the eastern part of the nebula shows a further protrusion in the form of an outer jet. While it is hard to distinguish between these scenarios, and disentangle the dominance of either the jets or torus in the present observation, the presence of torus/jets are strong indicators of $\zeta_{\rm PSR}$ and can constrain the geometry of the system as shown in Section~3.

To accurately determine the position of the point source (pulsar), we created a subpixel image of the same (at one-fifth of the ACIS pixel resolution) and applied the source detection algorithm {\it celldetect} in the soft energy band  of 0.5--2 keV where the pulsar is visible. The coordinates of the point source are RA(J2000)$=08^h55^m36.138^s$ Dec$=-46\deg44\arcm13.57$\arcs\ considering an error of 0.6\arcs\ at 90 $\%$ confidence level in absolute \chandra astrometry. This is consistent with the radio position of the pulsar. 

The net count rate from the point source after subtracting the nebular component is 0.0005 counts/s, and from the entire nebula is 0.02 counts/s. The details of background subtraction are discussed in the spectral analysis Section~\ref{sec-spec}. 
   
%%%%%%%%%%%
\section{X-ray spatial analysis}
\label{sec:xspec}
\begin{figure}
\centering
\includegraphics[scale=0.35]{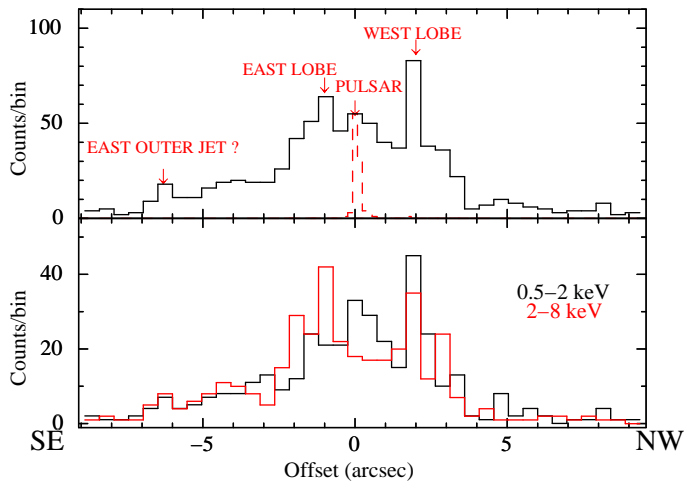}
\caption{The top panel shows the counts profile of the source extracted from the box region in Figure 1a (0.5--8 keV) compared with the 2-D PSF (0.5--8 keV, red dashed line). The bottom panel portrays the same profile in the two energy bands 0.5--2 keV and 2--8 keV. }
\label{radial}
\end{figure}

We performed detailed spatial analysis of the compact PWN surrounding \psr. To probe the structures of the nebula,  especially the axisymmetric features, we created a counts profile in the box region shown in Figure~\ref{images}a. We also performed morphological fitting of the PWN using the torus fitting model of \cite{ngromani2004,ngromani2008}, and determined its geometrical parameters like the torus radii, the position angle, and $\zeta_{\rm TORUS}$. 
%%%%%%%%%%%%%%%%%%
Figure~\ref{radial} shows the 0.5--8 keV counts profile decomposed into two energy bands. The profile was created in the box region of length 10\arcs\ shown in Figure~\ref{images}a, along the symmetry axis, averaged over a region (4.5\arcs\ in width) perpendicular to it. The box was centred with respect to the X-ray counterpart of the pulsar,  and we overlaid the point-spread function (PSF) in the figure. The PSF of the \obs was simulated using the \chandra ray tracer chaRT\footnote{http://cxc.harvard.edu/chart/} which simulates the High Resolution Mirror Assembly based on the energy spectrum of the source and the \obs exposure. The output of chaRT was modelled with the software MARX\footnote{http://cxc.harvard.edu/chart/threads/marx/} taking the instrumental effects and the EDSER subpixel algorithm into account to be consistent with the observational data. The best-fit spectrum of the point source (see Section~\ref{sec-spec}) was used. The profile clearly shows all the structures of the nebula including the two lobes symmetric about the pulsar. The western lobe has higher counts than the eastern one, although consistent within errors. The region west of the pulsar declines more steeply with radius, almost immediately after the lobed feature. The eastern part on the other hand has two additional bumps, the farthest one corresponding to a tail-like protrusion seen in the images, which might be the outer jet. The decomposed counts profile highlights the dominance of some features of the PWN in certain energy ranges. As seen from the images, and also from spectral analysis (later in Section~\ref{sec-spec}), the X-ray counterpart of the pulsar is clearly seen in the energy band of 0.5--2 keV, while it is less dominant above 2 keV.  The east lobe of the nebula shows an indication of being harder, being more dominant in the energy range of 2--8 keV.  The eastern outer jet has equal dominance in the soft and hard energy bands. 

%%%%%%%%%%%%%

%%%%%%%%%%%%%

%--------------------------------------------------------------------------------

%=============================================================================
%%%%%%%%%%%%%%%%%%
Section~2 discusses three possible scenarios for the morphology of the PWN around \psr. In two of the three scenarios where the lobes either correspond entirely to the arcs of the torus, or to one of the torus components along with the jets, it is possible to provide quantitative measurements of the PWN geometry by modelling the lobes as a double torus as in the pulsar wind torus fitting scheme of \cite{ngromani2008}. In the third case, identifying the lobes entirely as jets instead does not alter the geometry of the system as the torus axis is coincident with the spin axis of the pulsar along which jets are formed \citep{ngromani2008}. The parameters of the torus model are the torus axis position angle (PA) $\Psi$ (N-E)\footnote{Position angle 0\deg\, corresponds to North and 90\deg\, to East, observer angle $\zeta_{\rm TORUS}$ between the torus axis and the observer line of sight, radius $R$ and a finite thickness or "blur" $\delta$ of the cross section of the torus, and the post-shock flow velocity $\beta_{\rm shock}$. For the double torus there is an additional parameter \textit{d} for the separation between the two torii. It is symmetrically offset along the torus axis with the pulsar positioned in the middle. The torus axis is coincident with the pulsar spin axis, $\zeta_{\rm TORUS}=\zeta_{\rm PSR}$ and so $\zeta_{\rm TORUS}$ will be referred as $\zeta_{\rm PSR}$ in the rest of the paper.}

Spatial fitting was performed on the full resolution image, in a region of size 47\arcs\ $\times$ 45\arcs\ centred on the pulsar to constrain the background better. The PSF described in the previous section was loaded as a table model in {\it Sherpa} to model the pulsar emission. The remaining excess consisted of the background and the emission from the nebula. The background was modelled with a constant. The lobes of the nebula were modelled with the relativistic torus model of \cite{ngromani2008} as a double torus seperated by a distance \textit{d}. The background and the double-torus model was then convolved with the PSF. The $\delta$, if left free, has a tendency to grow and absorb the unmodelled larger-scale PWN components. As this parameter does not provide any physical information on the shock or the geometry of the system, it was frozen to 0.5\arcs\ provided by an initial visual estimate.
 
\begin{figure*}
\centering 
\includegraphics[scale=0.95]{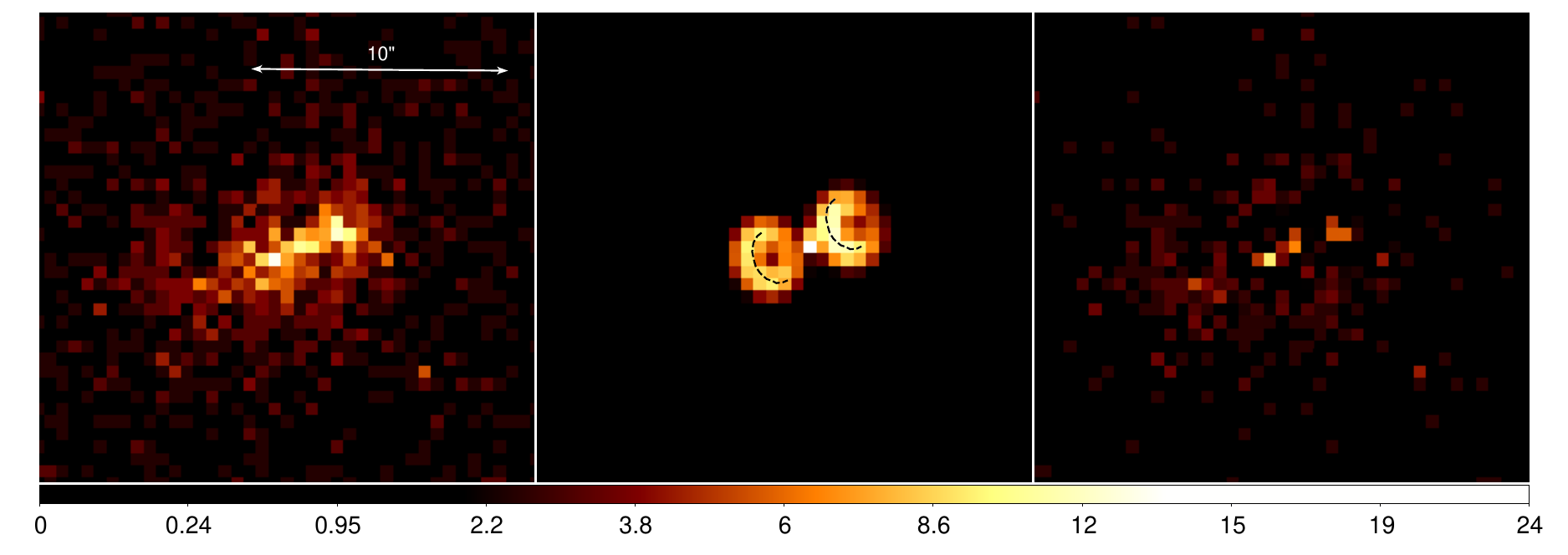} 
\caption{The first panel shows the data in the energy range of 0.5--8 keV. The second one shows the corresponding best-fit geometric model with a double torus \citep{ngromani2008}. The arcs of the torus are marked as black dashed lines to illustrate the geometry. The third  panel displays the residuals after the best-fit has been subtracted.The colour bar is in square root, and units are counts for all the images.}
\label{corr}
\end{figure*}

Figure~\ref{corr} shows the best-fit model along with the data used for fitting and the residuals. The best-fit parameters were determined using the C-statistic \citep{cash1979}\footnote{The Cash statistic is a likelihood function defined using Poisson statistics. Probability P$_{x,y}$ of observing d$_{x,y}$ counts out of the expected c$_{x,y}$ counts in pixel (x,y) is $P_{x,y}=\frac{c_{x,y}e^{-c_{x,y}}}{d_{x,y}!}$ . It may be used regardless of the number of counts in each pixel and passes to $\chi^{2}$ statistics at c$_{x,y} \ge 20$. The best-fit parameters are determined by the negative logarithm of the likelihood function summed over all image pixels.}. The best-fit parameters are listed in Table~\ref{tabsherpa}. The 1$\sigma$ confidence intervals are quoted here. The statistical errors were estimated by performing Monte Carlo Poisson realizations of the best-fit model as recommended in \cite{ngromani2008}. To characterize the {\bf uncertainty on the estimates associated with the unmodelled components in the fit}, especially the eastern tail region, we blanked out the region in the fit and noted the change in parameters. Considering that the {\bf two} errors add in quadrature, we find that the pulsar line of sight $\zeta_{\rm PSR}$ is inclined at $32.5\dg\pm 4.3\dg$ with an upper limit of $\zeta_{\rm PSR}$ < $37\dg$. 

%%%%%%%%%%%

%\begin{figure*}
%\centering 
%\includegraphics[scale=0.6]{figs/fit-psf.ps} 
%\caption{First {\bf leftmost} shows the data; {\bf the middle panel} the best-fit model; and {\bf rightmost panel} the residuals after the {\bf best fit model has been subtracted. The model was convolved with PSF. The pulsar was {\bf modelled} with the PSF.}}
%\label{corr}
%\end{figure*}

%%%%%%%%TABLE1%%%%%%%%
\begin{table}
\centering
\caption{Final morphological best fit with a double-torus model. The model was convolved with the PSF as described in the text.}
\begin{tabular}{c c c}
\hline

Parameter  & Value & Units\\
 \hline         
$\Psi^{a}$  & $114.4\pm 2.3 \pm 1.1$ & $\dg$ \\
$\zeta_{\rm PSR}$ & $32.5 \pm 4.0 \pm 1.7$ & $\dg$ \\
R & $2.1\pm 0.06 \pm 0.08$ & \arcs\ \\
$\beta_{\rm shock}$ & $0.41\pm  0.06 \pm 0.06$ & --\\
\textit{d} &  $3.6\pm0.70 \pm 0.30$ & \arcs\ \\
Background & $0.46\pm0.02$ & counts/pixel \\
\hline    
\end{tabular}
\\
$^{a}$ For all the parameters except the background counts/pixel, the first error denotes the statistical error, and the second the {\bf uncertainty on the estimates associated with the unmodelled components in the fit}.\\
\label{tabsherpa}
\end{table}

%%%%% Spectral analysis section %%%%%%%%%%%%%
\section{X-ray spectral analysis}
\label{sec-spec}
\begin{figure*}
\centering
\includegraphics[scale=0.63,angle=0]{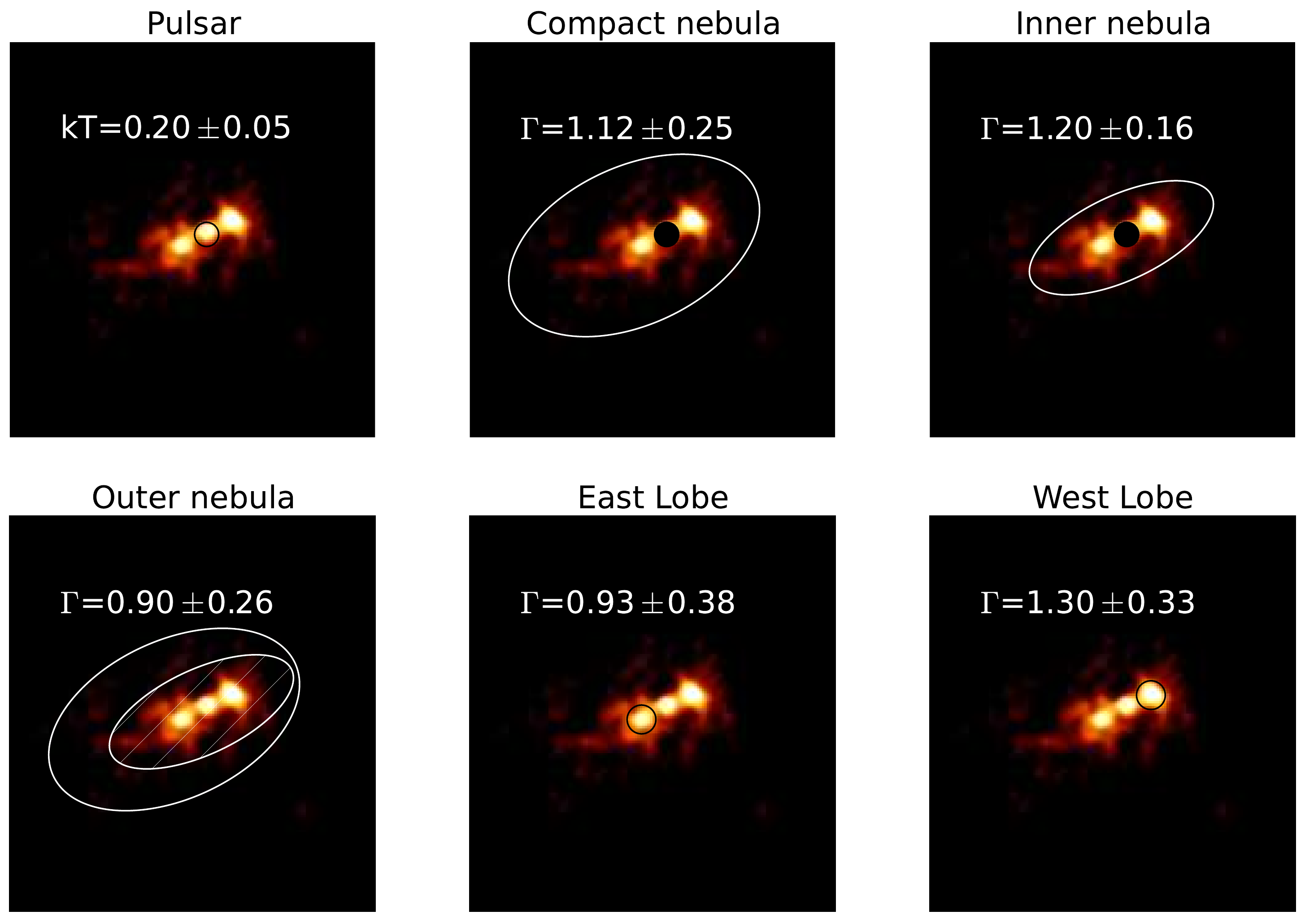}
%\subfigure[]{\includegraphics[height=0.45\textwidth,angle=-90]{figs/fig3b_total_neb.ps}}
 \caption{The sub-figures show the \chandra image (0.5--8 keV) of the PWN around \psr~along with the regions used for spectral extraction shown in white. The value of the $\Gamma$ (or $kT$) obtained from spectral fitting is also indicated in each panel.}
\label{spec-1}
\end{figure*}

We have performed a detailed spectral analysis of the X-ray counterpart of \psr~and the surrounding compact PWN. To look for changes in the spectral parameters in different parts of the nebula we also extracted spectra from different regions by dividing the PWN into an inner and outer annular part, and have examined the spectra of the east and west lobes separately. The analysis was performed in the energy range of 0.5--8 keV. The C-statistic was used for spectral fitting and errors were estimated at a 90$\%$ confidence interval. The regions used for spectral extraction are shown in Fig.~\ref{spec-1}. For the pulsar, a circular region of radius 0.7\arcs\ was extracted, centred on the best-fit coordinates of the source. In the case of the compact PWN, the outer radius was optimized from the spatial fitting, with the region corresponding to the pulsar excised. 
 For all the regions, the background spectrum was extracted from an annular region with inner and outer radii of 75\arcs\ and 100\arcs\ (extent of the diffuse nebula seen with {\it XMM}: see Fig.~\ref{images}d). This ensured the background was free from the fainter diffuse nebular emission that was seen with the \xmmn observations. In the case of the pulsar, apart from the annular background spectrum, the astrophysical background due to the nebula was considered, and was modelled with the best-fit spectrum of the entire PWN, scaled with the ratio of the two extraction regions.  We also checked for an additional astrophysical background contribution from the pulsar in the nebular region, and found it to be a negligible fraction of the total background. 
 
\begin{figure*}
\centering
\subfigure[]{\includegraphics[height=0.45\textwidth]{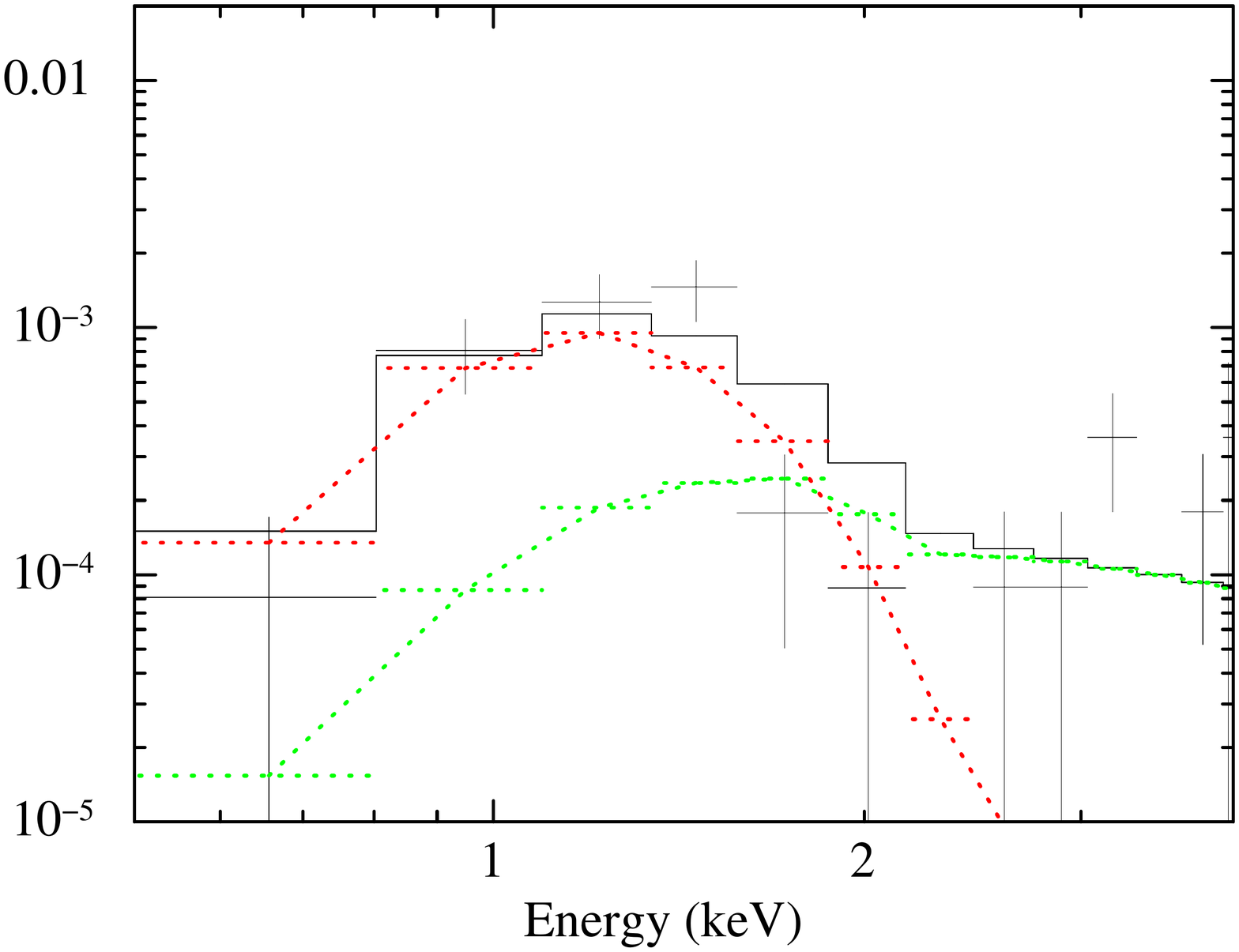}}
%\subfigure[]{\includegraphics[height=0.45\textwidth,angle=-90]{figs/fig3a-pulsar.ps}}
\subfigure[]{\includegraphics[height=0.45\textwidth]{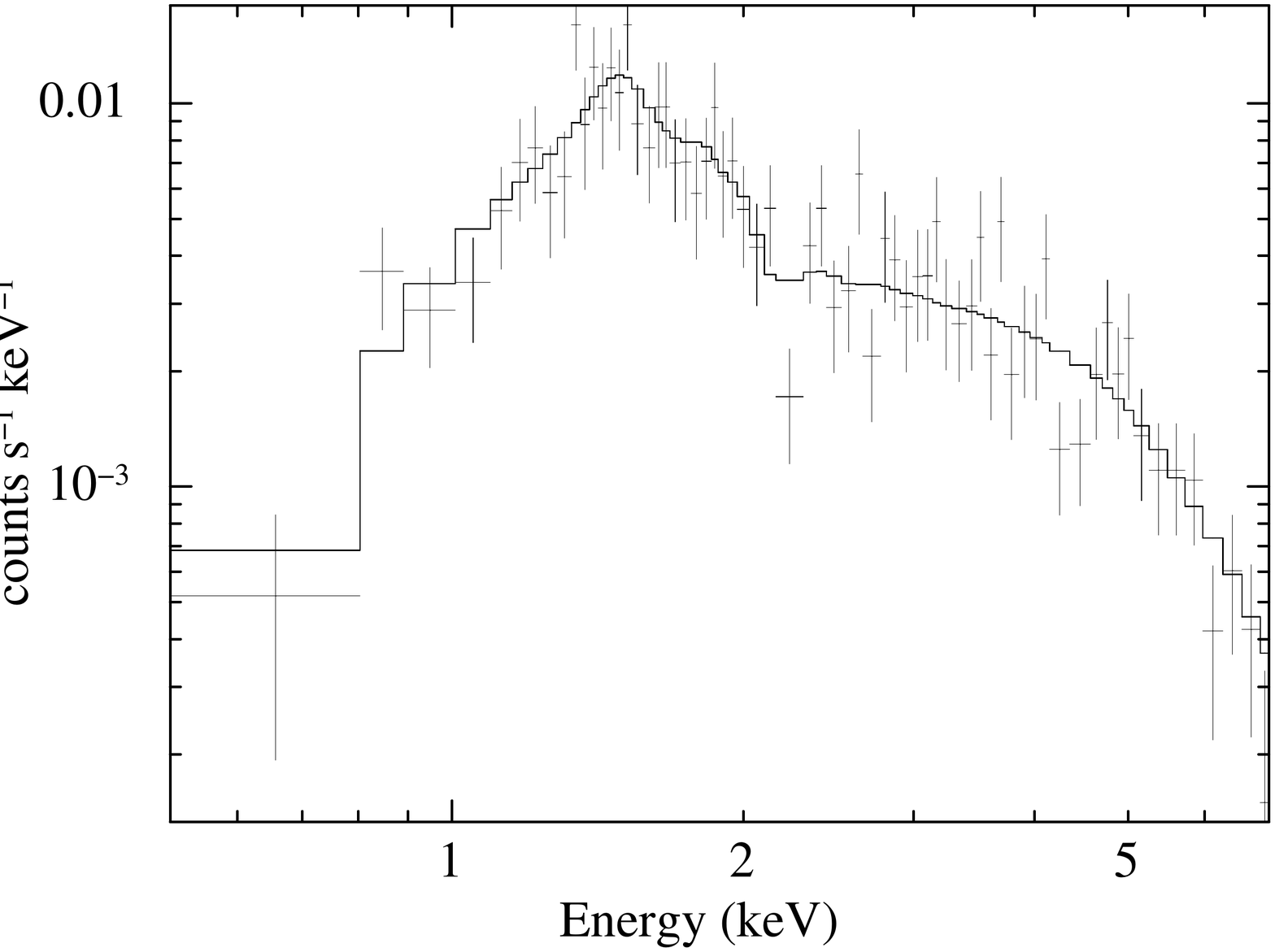}}
 \caption{The panels show the ACIS-S spectra for the X-ray counterpart of the pulsar \textit{(Panel~(a))} and the entire PWN \textit{(Panel~(b))}.
The solid lines correspond to the respective best-fit spectral models (absorbed blackbody and absorbed  power law, respectively). In the left  panel, the pulsar blackbody component (red) and the nebular contribution in the same region (green) are shown. The unbinned spectrum fitted using C-statistics have been rebinned in the plot for visual clarity.}
\label{spec-2}
\end{figure*}

We used an absorbed  power law as the spectral model for the compact PWN and all the decomposed regions marked in Fig.~\ref{spec-1}. In order to account for the photoelectric absorption by the interstellar gas along the line of sight, a free absorption ({\it XSPEC} model {\it tbabs}) component was used. This component was determined from the fit of the entire nebula and frozen henceforth for all other regions as it is not expected to vary locally. The value obtained  with \chandra is  $N_{\rm H} = 0.70^{0.25}_{-0.20}\times10^{21}$~cm$^{-2}$, which is consistent with that obtained using {\it XMM-Newton}, derived from a 15\arcs\  radius region ($N_{\rm H}$ = 0.64$^{0.13}_{-0.11}\times10^{21}$~cm$^{-2}$). 
The spectrum of the PWN is hard with a power-law index $\Gamma$ of $1.12\pm0.25$. The unabsorbed bolometric luminosity L$_{\rm eff}$ (0.5--8 keV) is 3.3$\times10^{31}\ergsec$ assuming a distance of $d\sim900$~pc. Comparing the spectra of the inner and outer nebula, we did not detect any steepening of the power-law spectral index $\Gamma$ which could have been indicative of synchrotron cooling. In the case of the east and west lobe, while there is an indication from the counts profile (Fig.~\ref{radial}) that the eastern lobe has a harder spectrum than the western region (higher counts in 2--8 keV band in the east lobe compared to 0.5--2 keV), we do not find a significant difference in the spectral indices, given the current level of statistics. The pulsar spectrum is faint and soft with the statistics deteriorating drastically beyond 5 keV. The spectrum was fitted with an absorbed blackbody as well as a neutron star hydrogen atmosphere model \citep[{\it nsa};][]{2002yCat..33861001G,zalvin2009}. Both models provide acceptable fits to the data with comparable C-stat values. The fit with the {\it nsa} model shows an effective temperature T$_{\rm eff}$ of $2.2 \times 10^{6}$ K and an unabsorbed bolometric luminosity $L_{\rm eff}$ (0.5--8 keV) of $1.4 \times10^{30}\ergsec$ assuming a distance of $d=900$~pc. The blackbody fit of the thermal component gives a similar temperature $T_{\rm BB}$ of $2.0  \times10^{6}$ K  and  $L_{\rm eff}$ of $1.3 \times10^{30}\ergsec$. Both indicate an effective radius $R_{\rm eff}$ of $\sim 1.5$~km suggesting that the emission originates from hot spots on the poles of the neutron star, rather than being an emission from the entire surface. An additional power-law component (related to potential non-thermal emission) was not required for the fit. 
The pulsar and the compact PWN spectrum along with its best-fit model and residuals are shown in Fig.~\ref{spec-2}a and ~\ref{spec-2}b, and the best-fit parameters for all the regions along with the C-statistic values are given in Table~\ref{table-specfit}.  
%%%%%%%%%%%%%%%%%%%

 %%%%%%%%%%%%%%%%%
%%%%%%%%%%%%%%%%%%%

%%%%%%%%%%%%%%%%%

%%%%%%%%%%%%%%%%%%%%%%%
\begin{table*}
\caption{Parameters of the best-fit spectral model to the putative pulsar and nebula with background,
without diffuse emission. Errors are quoted at 90$\%$ confidence. Absorption is fixed in all the fits, except in the case of the total nebula.}
\centering
\begin{tabular}{lccccccc}
\hline
Region & Absorption & Photon index & $kT_{\rm BB}^{\inf}$ & $kT_{\rm eff}^{\inf}$ & $F_{x}$$^{a}$ &  $F_{x}$$^{a}$  & C-stat/$N_{\rm dof}$\\
 -- & $10^{21}$cm$^{-2}$  &  $\Gamma$  & keV &  keV&  0.5--8 keV &  2--10 keV \\
\hline
Pulsar &0.70(f) & --& $0.20\pm0.05$ & $0.22\pm0.05$ & ($2.48\pm1.04$)$\times10^{-15}$ & -- & 248.48/442 \\
Compact nebula & $0.70^{+0.25}_{-0.20}$ & $1.12\pm0.25$ & -- & -- & ($2.80\pm0.24$)$\times10^{-13}$ & ($3.2\pm0.42$)$\times10^{-13}$ & 466.50/440\\
Inner nebula & 0.70 (f) & $1.20\pm0.16$  & --& -- & ($1.94\pm0.40$)$\times10^{-13}$ & ($2.17\pm0.32$)$\times10^{-13}$\ & 434.86/440\\
Outer nebula & 0.70 (f) & $0.90^{+0.28}_{-0.24}$ & -- & -- & ($8.87\pm1.3$)$\times10^{-14}$ & ($1.07\pm0.21$)$\times10^{-13}$ & 369.53/440\\
West lobe & 0.70 (f) & $1.30\pm0.33$  & -- & -- & ($3.24\pm0.72$)$\times10^{-14}$ & ($3.55\pm1.02$)$\times10^{-14}$ & 255.71/442\\
East lobe & 0.70 (f) & $0.93\pm0.38$ & -- & -- & ($3.15\pm0.78$)$\times10^{-14}$ & ($3.86\pm1.20$)$\times10^{-14}$ & 234.26/442\\

\hline 
\label{table-specfit}   
\end{tabular}
\\
$^{a}$Observed flux in units of $\ergcms$ \\
(f) Parameter fixed for consistency between fit regions.  \\
Instrumental line at $1.49 \pm 0.11$ keV sigma fixed.

\end{table*}

%%%%%%

\section{Constraints from geometric light curve modelling}
The fact that \psr~is radio-loud but $\gamma$-quiet (despite its large pseudo luminosity $\dot{E}/d^2$) may be used to derive constraints on the pulsar geometry. The radio peak multiplicity and width may also aid in obtaining geometric constraints. One can compare the predictions of geometric light curve models with the observed radio profile to constrain both $\alpha$ and $\zeta_{\rm PSR}$. We used the TPC model \citep{Dyks03} and a geometric version of the OG model \citep{CHR86a,Romani96}, in conjuction with a semi-empirical hollow-cone radio model \citep{Gonthier04,Harding07_Geminga}. The latter model is based on work by \citet{Rankin93,ACC02}, with the radio emission altitude expression as in \citet{Kijak03}. For more details, see \citet{Venter09}. 

We generated an `atlas' of $\gamma$-ray and radio light curves on an ($\alpha$,$\zeta_{\rm PSR}$) grid, at a 5$^{\circ}$ resolution. In the next two sections, we attempt to provide conservative limits on $\alpha$ and $\zeta_{\rm PSR}$ based on the qualitative features of the observed radio profile (peak multiplicity, duty cycle) and the non-detection of $\gamma$-ray pulsations. Additionally, we perform a $\chi^2$ fit using the observed radio profile to further constrain $\alpha$ and $\zeta_{\rm PSR}$, in a subsequent section.

\subsection{Radio visibility and peak multiplicity}
For the radio models, we used period of $P = 65$~ms. This important parameter sets the radio beam width (assuming emission from a single altitude for simplicity), which scales as $P^{-1/2}$ (as does the polar cap opening angle $\theta_{\rm PC}$, with which it is closely associated). The beam width, in conjunction with the observer angle $\zeta_{\rm PSR}$, determine the radio visibility of the pulsar. Conversely, the $\gamma$-ray emission originates from a range of altitudes and undergo beaming effects, so the caustic $\gamma$-ray beam shape in this case is relatively independent of the period (unless the observer's line of sight crosses the non-emitting polar cap on the stellar surface when $\alpha\approx\zeta_{\rm PSR}$, which will cause a dip in the $\gamma$-ray light curves). We set the radio frequency to $\nu=1400$~MHz (same as for the observed profile). The observer orientation with respect to the radio beam provides three regimes in terms of radio pulse shape predictions. For an impact angle $\beta \equiv \zeta_{\rm PSR}-\alpha\sim 0\dg$, a double-peaked structure will be seen. For larger values of $|\beta|$, a single peak will be visible, since the observer skims the edge of the hollow cone. For even larger values of $|\beta|$, the cone will be completely missed, and the pulsar will be radio-quiet. Additionally, for small $\alpha$, the duty cycle is large, and for large $\alpha$ and small enough $|\beta|$, one starts to see radio beams from both magnetic poles (leading to two profiles with about an 0.5 difference in normalized phase). Therefore, the fact that the observed radio peak of \psr~is single provides a lower limit $|\beta|\gtrsim10\dg-15\dg$ (the lower value is for low values of $\alpha$). On the other hand, radio visibility provides an upper limit on $|\beta|$: from the radio model we can derive the constraint $|\beta|\lesssim 30\dg$. This result is somewhat dependent on the assumed emission profile of the cone. Currently, we use an offset-Gaussian flux profile that peaks at a co-latitude $\bar{\theta}$. Since we assume emission from a single height, this maximum relative flux is independent of $\alpha$ and $\zeta_{\rm PSR}$. The Gaussian profile has tails reaching beyond the last open $B$-field lines (which are tangent to the light cylinder at a cylindrical radius $r=cP/2\pi$, where the corotation speed equals the speed of light), into the corotating region. If one wants to restrict the emission by not allowing any radio flux from this region (i.e., terminating the tails at $\theta=\theta_{\rm PC} \approx \sqrt{2\pi R/Pc}$), this would lead to smaller radio beams (by a few degrees) and therefore to tighter constraints on $|\beta|$. However, given the inherent uncertainties in the beam geometry, it might not be wise to pursue such stricter constraints on $|\beta|$ that depend on the detailed beam structure. We can therefore summarize our constraint from this section as $10\dg\lesssim|\beta|\lesssim 30\dg$. This constraint is also borne out by the $\chi^2$ fit to the data performed below.

\subsection{$\gamma$-ray invisibility (non-detection)}
\begin{figure*}
\includegraphics[width=0.5\textwidth]{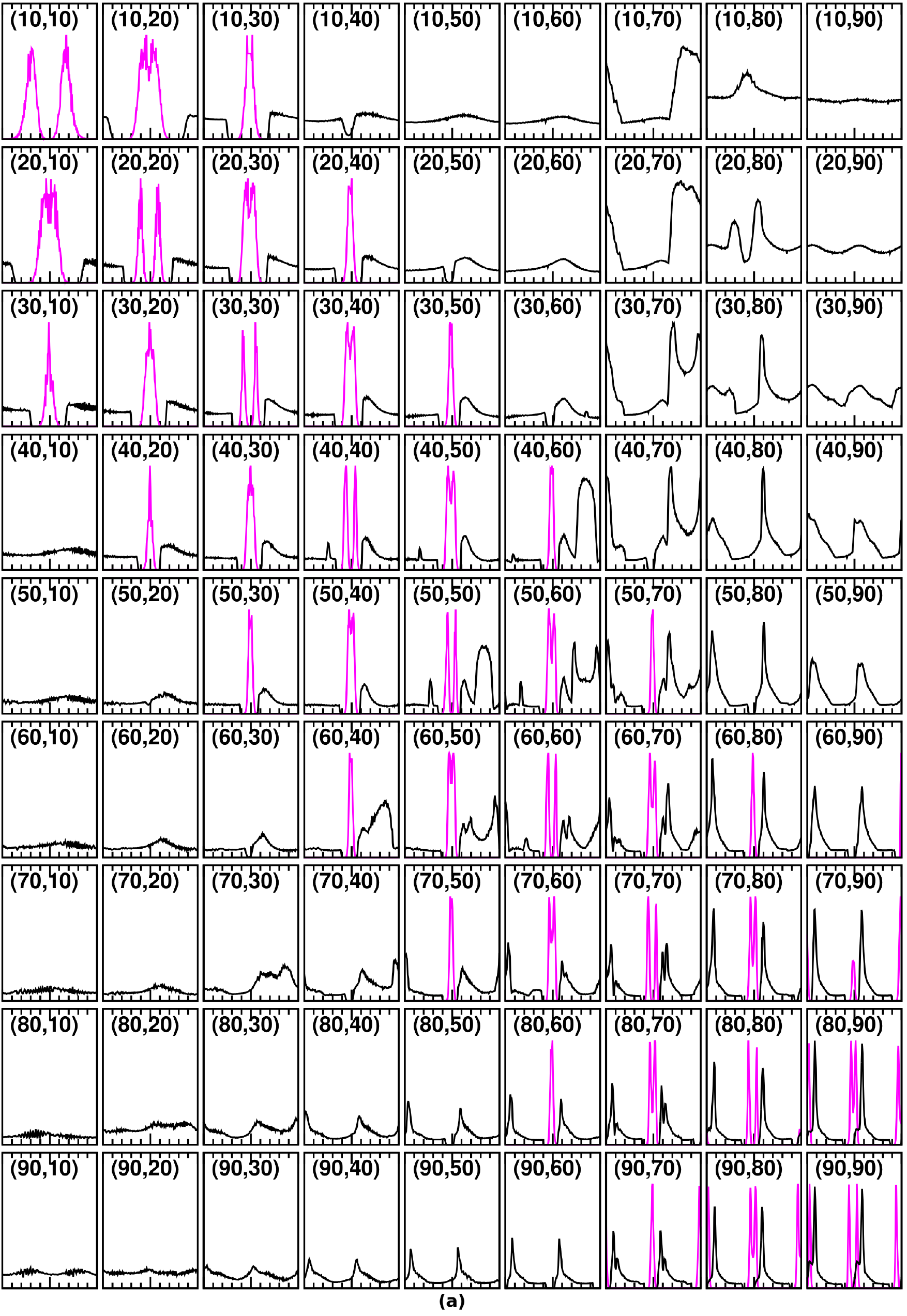} 
\includegraphics[width=0.5\textwidth]{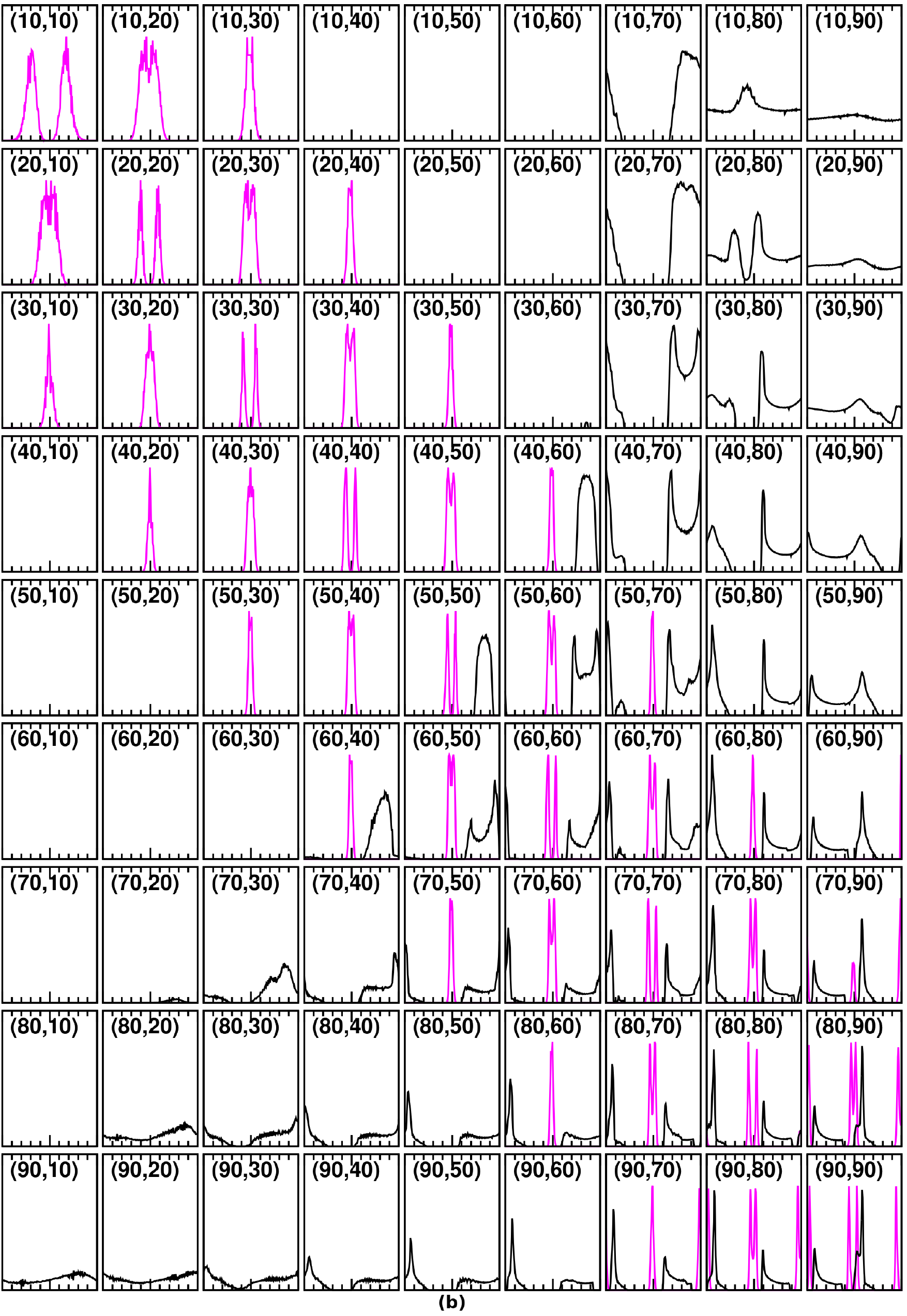} 
%\caption{{ Same as textit{(Panel~(a))} , but for an OG model.}}
\caption{Plot of predicted TPC (\textit{Panel (a)}) and OG (\textit{Panel (b)}) $\gamma$-ray (black) and radio (magenta) light curves on a $(\alpha,\zeta_{\rm PSR})$ grid. The $\gamma$-ray intensity has been normalized globally so that the maximum (typically at $\alpha=\zeta_{\rm PSR}=90\dg$) is unity and the flux of the other combinations of $\alpha$ and $\zeta_{\rm PSR}$ relative to this maximum is clear. The $x$-axis denotes normalized from -0.5 to 0.5.  A phase offset of 0.5 has been artificially imposed to centre the radio pulses for visual clarity. The coordinate pair in each panel refers to the value of $\alpha$ and $\zeta_{\rm PSR}$ in each case.}\label{fig:TPC}
\end{figure*}

The pulsar geometry is much better constrained if radio and $\gamma$-ray data (or a non-detection in the latter case) are used simultaneously \citep{Pierbattista15}. In our case, the lack of $\gamma$-ray visibility leads to further constraints on $\alpha$ and $\zeta_{\rm PSR}$, within the context of the TPC and OG models. In Figure~\ref{fig:TPC}, we have plotted $\gamma$-ray (black) and radio (magenta) light curves on an $(\alpha,\zeta_{\rm PSR})$ grid at a resolution of 10$^{\circ}$. We have normalized the $\gamma$-ray intensity so that the maximum (typically at $\alpha=\zeta_{\rm PSR}=90\dg$) is unity and the flux of the other combinations of $\alpha$ and $\zeta_{\rm PSR}$ relative to this maximum is clear\footnote{This relative flux is not seen for the radio pulses, since the radio code predicts the same maximum conal flux from a single altitude, based on $P$, $\dot{P}$, and $\nu$ at $\theta = \bar{\theta}$, independent of $\alpha$ and $\zeta_{\rm PSR}$, as mentioned earlier.}. The geometric models cannot predict absolute\footnote{Absolute predictions of flux, and therefore relative $\gamma$-ray vs.\ radio flux, are only possible when using full radiative models, which is outside the scope of the current paper.} $\gamma$-ray fluxes, neither can they give an indication of the $\gamma$-ray flux relative to the radio flux at a particular viewing geometry, so one cannot use a sensitivity-limit argument to constrain $\alpha$ and $\zeta_{\rm PSR}$. Rather, one can only use visibility (and radio pulse shape) arguments to constrain these angles. 

We note that we have followed \citet{Venter09} and assumed a gap width of 0.05$\theta_{\rm PC}$ for both the TPC and OG models. It is true that in physical pulsar models, the gap width sensitively depends on the pulsar period and magnetic field. For example, \citet{Muslimov2003} give expressions for the slot gap width as a function of $P$ and $B$, and the width scales linearly with $P$. When using geometrical models, however, the usual approach is to assume a reasonable value for this width which results in good light curve fits. Our choice of 5\% is also supported by the results of \citet{Johnson2014} who fit the light curves of 40 $\gamma$-ray millisecond pulsars, finding that their best-fit gap width never exceeded 10\%, but is usually smaller. Further justification comes from Figure~2 of \citet{Muslimov2003}. While it is true that the gap width influences the shape of the resulting $\gamma$-ray light curves, the actual $\gamma$-ray pulse shape is of less importance in our case due to the following reasons: (i) Observational data imply that the gaps must be narrow in order to reproduce the observed shapes; (ii) We are only using the geometrical gap models to decide for which parameter ranges ($\alpha$ and $\zeta_{\rm PSR}$) the $\gamma$-ray pulse should be visible. Therefore, the width of the pulse is less important than the relative intensity at different $(\alpha,\zeta_{\rm PSR})$, and different gap width choices should not alter our main conclusions. As for the magnetic field structure, we have assumed the vacuum retarded dipole \citep{Deutsch55,Cheng00,Dyks04}, which has been standard in recent years, allowing other authors to compare their results to ours.

For the TPC model, one can see that the $\gamma$-ray light curves are visible at almost all angles (except $\alpha=\zeta_{\rm PSR}\sim10\dg$). The low-level $\gamma$-ray emission reflects the emission outside of the main caustic beam. However, the relative flux increases dramatically when the observer samples the bright caustic (e.g., compare $(\alpha,\zeta_{\rm PSR})=(10\dg,60\dg)$ and $(\alpha,\zeta_{\rm PSR})=(10\dg,70\dg)$). Given the $\gamma$-quiet and single visible radio pulse constraints, this would imply both $\alpha$ and $\zeta_{\rm PSR} \lesssim 50\dg$, assuming that the low level of $\gamma$-ray flux (and broad $\gamma$-ray peaks) would not be detectable. The case for the OG model is perhaps a bit clearer. Since no $\gamma$-ray emission is generated below the null charge surface where the Goldreich-Julian charge density is zero \citep{GJ69}, $\gamma$ rays are only visible for $\zeta_{\rm PSR} \gtrsim50\dg$, and for large $\alpha$. For example, at $(\alpha,\zeta_{\rm PSR}) \sim (50\dg,50\dg)$, one starts to sample the main caustic beam. (For larger $\zeta_{\rm PSR}$, one samples the caustic twice during one rotation, leading to double peaks.) Since the radio model is the same for both the TPC and OG cases, one derives the same constraints on $\alpha$ and $\zeta_{\rm PSR}$ for both $\gamma$-ray models. At a finer resolution, one finds that the constraint may be written as $\alpha \lesssim 55\dg$ and $\zeta_{\rm PSR} \lesssim 55\dg$. It would be difficult to pick a best model between the OG and TPC in this case, since this would depend on whether the low-level TPC emission is deemed detectable or not. This is difficult to decide given the lack of absolute flux predictions.

\subsection{Radio profile fit}
In this section, we constrain $\alpha$ and $\zeta_{\rm PSR}$ by fitting the observed radio profile from \cite{2003MNRAS.342.1299K} using our radio model and a $\chi^2$ minimization procedure. We are only interested in reproducing the pulse shape, so we normalize the flux maximum to unity. We estimate the error on the data by calculating the standard deviation of the flux for the phase range $\phi<0.4$ (this range was chosen as the `off-peak region'), yielding $\sigma_{\rm data} \sim 0.06$. The number of degrees of freedom is $N_{\rm dof} = 255$. We used $\chi^2 = \sum_i[(Y_{\rm model}-Y_{\rm data})/\sigma_{\rm data}]^2$ for $i=1,\dots,N_{\rm bins}$. The radio model was run for $\alpha \in [0\dg,90\dg]$ and $\zeta_{\rm PSR} \in [0\dg,90\dg]$ at a $1\dg$ resolution, for $P=65$~ms and $\nu=1400$~MHz, and initially for 720 phase bins. In order to perform a $\chi^2$ fit, we rebinned (smoothed) the model to have the same number of bins as that of the data ($N_{\rm bins} = 256$) using a Gaussian Kernel Density Estimator (KDE) with step size $h=1/N_{\rm bins}$ \citep{Rosenblatt56,Parzen62}. We furthermore minimized $\chi^2(\alpha,\zeta_{\rm PSR})$ for each fixed value of $\alpha$ and $\zeta_{\rm PSR}$ over a phase shift parameter $d\phi \in [0,0.99]$ with a resolution of $0.01$. This is to allow alignment of the model and observed light curves, circumventing any issues that may stem from the uncertain definition of $\phi=0$ \citep[cf.][]{Johnson2014}. The background level was set to zero, and the model curve was treated as being cyclical. 

\begin{figure}
\hspace{-0.2cm}
\centering 
\includegraphics[scale=0.55]{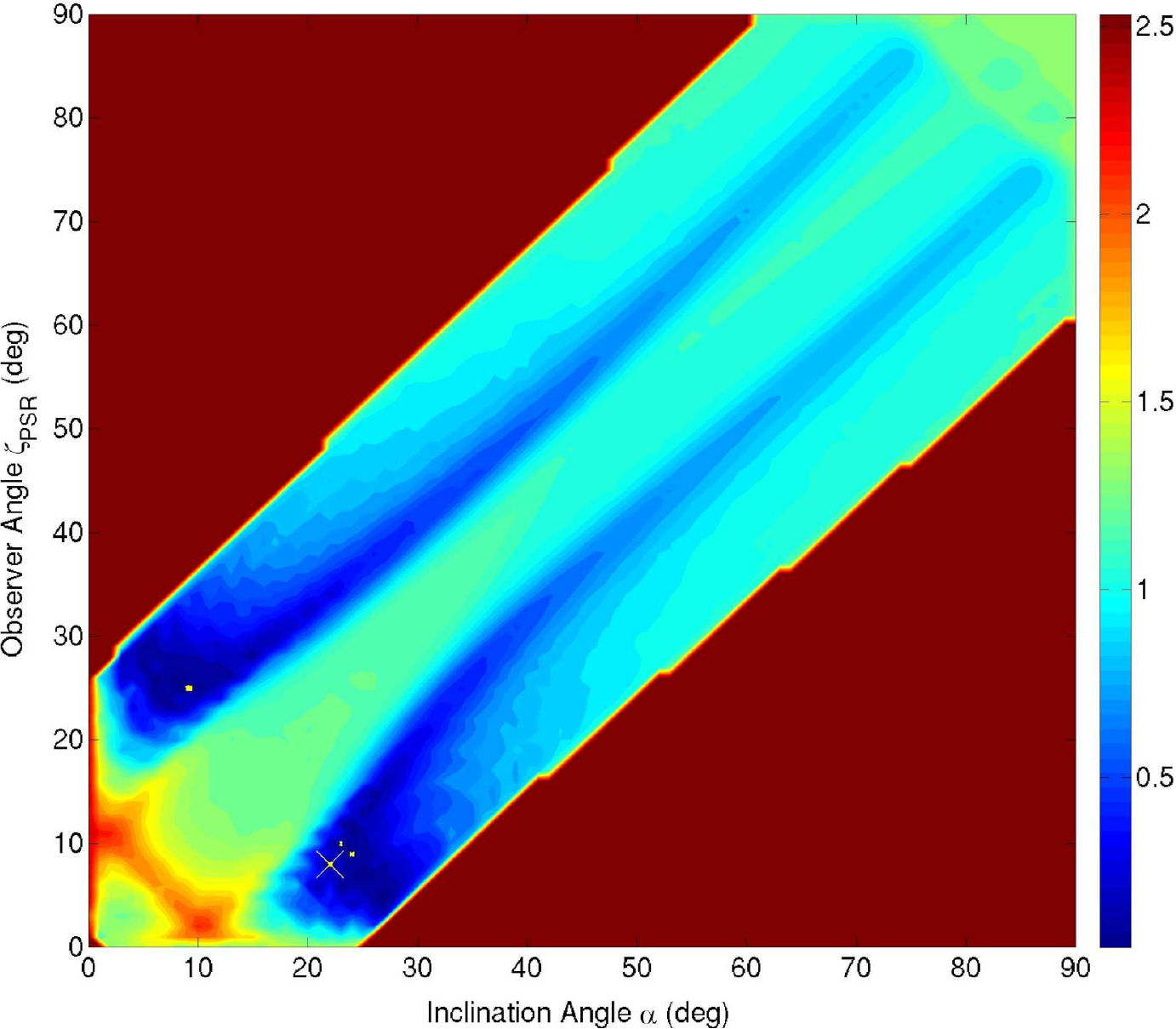} 
\caption{The $\log_{10}\chi^2/N_{\rm dof}(\alpha,\zeta_{\rm PSR})$ map with the best fit indicated by a yellow cross. The $1\sigma$, $2\sigma$, and $3\sigma$ contours are very small and are indicated by yellow lines.\label{fig:chi2}}
\end{figure}

\begin{figure}
\hspace{-0.2cm}
\centering 
\includegraphics[scale=0.55]{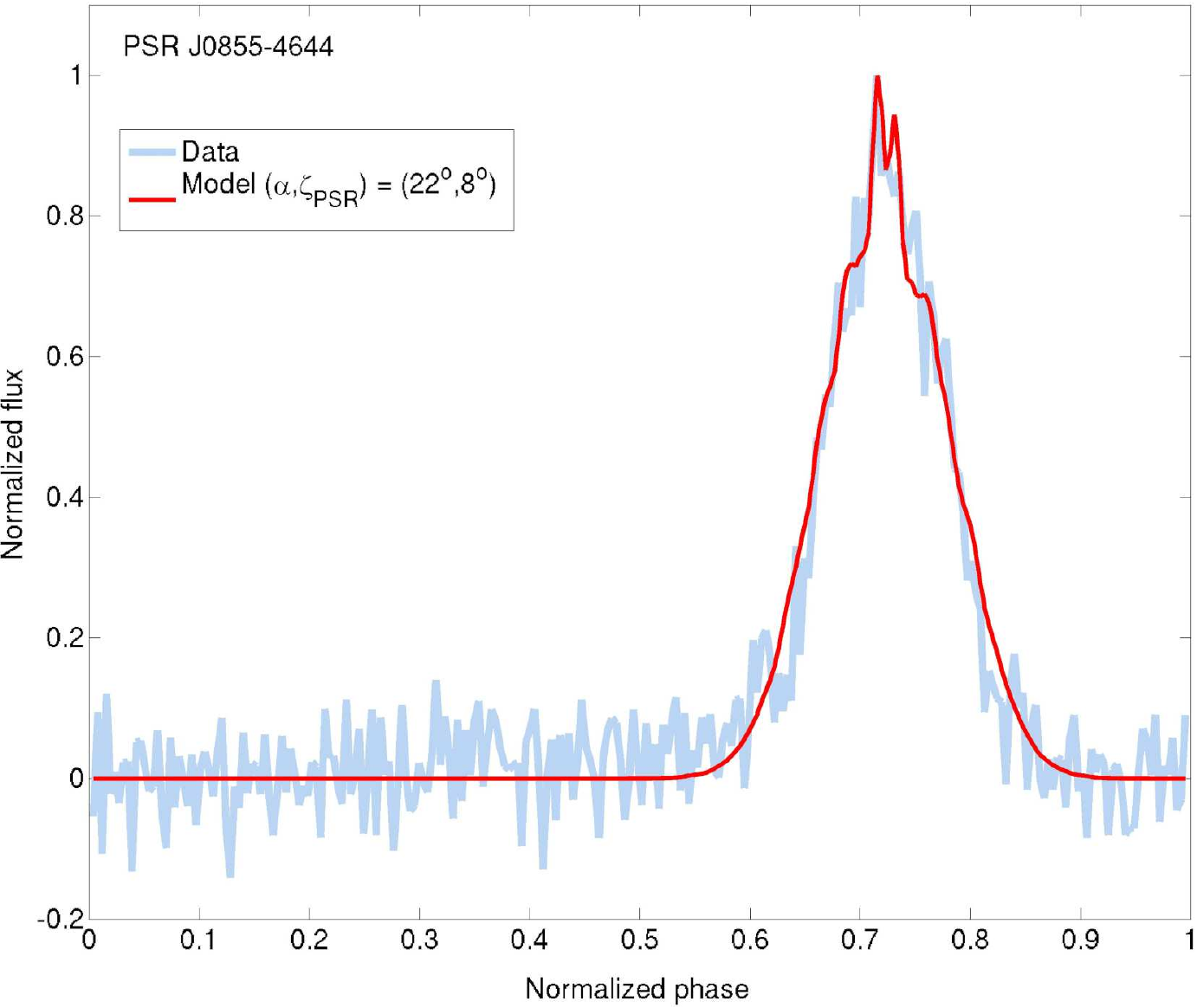} 
\caption{The best-fit radio light curves overplotted on the data.\label{fig:bestfitLC}}
\end{figure}

Figure~\ref{fig:chi2} shows the $\log_{10}[\chi^2(\alpha,\zeta_{\rm PSR})]$ map. The dark red colour indicates where the radio is invisible ($|\beta|\gtrsim30\dg$ as mentioned above). As one moves inward from larger to smaller $|\beta|$ values, the edge of the cone is clipped and the pulse is too narrow (cyan colour, $|\beta|\sim25\dg$); at $|\beta|\sim10\dg-15\dg$ a wider single peak is obtained that reasonably fits the data (blue colour), and at even smaller $|\beta|\lesssim10\dg$, a double peak is obtained (greenish colour). At small $\alpha$ and $|\beta|$ (e.g., $\alpha\sim15\dg$ and $\beta\sim5\dg$), the predicted single peak is very wide (yellow colour), and at $\alpha\gtrsim0\dg$ and $|\beta|\sim5\dg$, the edge of the cone is clipped for most of the rotation, so that the light curve is nearly flat, or very broad, with a very large duty cycle (light red colour). The best fit is found at $(\alpha,\zeta_{\rm PSR}) = (22\dg, 8\dg)$ for $d\phi = 0.73$, giving $\chi^2/N_{\rm dof} = 285/255$. The sigma contours (yellow lines) are calculated using $\log_{10}[\chi^2_{\rm min}+\sigma_i]$, with $\sigma_1 = 2.3$, $\sigma_2 = 6.18$, and $\sigma_3 = 11.83$ \citep{Lampton76}. They are quite small (and barely visible on the $\chi^2$ contour plot) and the $1\sigma$ contours imply strong constraints on the best-fit values of $\alpha$ and $\zeta_{\rm PSR}$. However, we note that there is an alternative fit at $(\alpha,\zeta_{\rm PSR}) = (9\dg, 25\dg)$ for $d\phi = 0.74$, giving $\chi^2/N_{\rm dof} = 294/255$, which lies within $3\sigma$ of the best fit. Other fits within $3\sigma$ are $(\alpha,\zeta_{\rm PSR}) = (23\dg, 10\dg)$ and $(\alpha,\zeta_{\rm PSR}) = (24\dg, 9\dg)$. These best-fit values satisfy the constraints derived in the two preceding sections. Figure~\ref{fig:bestfitLC} shows the best-fit radio light curve overplotted on the data.

%%%%%%%%

%%%%%%%%

%%%%%%%%

%%%%%%%%

% \begin{figure}
% \hspace{-0.2cm}
% \centering 
% \includegraphics[scale=0.27]{figs/angle1.eps} 
% \includegraphics[scale=0.27]{figs/angle2.eps} 
% \includegraphics[scale=0.27]{figs/angle3-new.eps} \label{fig:pp}
% \caption{Example plots of predicted radio (green) light curves with different $(\alpha,\zeta_{\rm PSR})$ marked in plot. The angles are denoted in degrees. Overlayed (blue) is the observed radio profile adapted from \cite{2003MNRAS.342.1299K}. The plots in the first two rows indicate the subspace of $(\alpha,\zeta_{\rm PSR})$ for which the radio profile matches most closely to the model. In the third row, some profiles which do not match the radio data are shown. In the leftmost plot of the third row, although $(\alpha,\zeta_{\rm PSR})$ lies in the expected range, $\beta$=10 is at the lower limit and produces a wider profile than observed.  In the middle one, $(\alpha, \zeta_{\rm PSR},)$ lies in the expected range as before, but $\beta$=25 is at the upper limit and produces a narrower profile. The rightmost plot demonstrates a scenario where $\alpha$ or $\zeta_{\rm PSR}$ is at the upper limit and produces a narrower profile than observed. The predicted light curves have been smoothed for better comparison with the data.}
%
%\end{figure}
%%%%%%%%%%%%%%
\section{Discussion}
Thanks to a dedicated \chandra observation, the compact PWN around the fast-spinning energetic pulsar \psr~has been resolved. The presence of two lobes, axisymmetric about a fainter pulsar counterpart, can have contributions from either the torus or jet structures or both of them, with the dominance of one component over the other indistinguishable by the present observation. Using the double torus model of \citet{ngromani2008} to spatially fit the lobes captures the characteristic features of the PWN. Most importantly, it provides a robust estimate of $\zeta_{\rm PSR}$ agreeable with constraints derived from geometric modelling of the radio pulse profile and non-detection of $\gamma$-ray pulsations from the pulsar.

\subsection{X-ray pulsar}
The X-ray counterpart of \psr~is faint and soft with a blackbody temperature of $0.2\pm0.05$ keV and  $L_{\rm eff}$ of $1.3\times10^{30}\ergsec$. The blackbody is emitted from a region with $R_{\rm eff}\sim1.5$~km, which indicates a polar cap nature of this emission rather than from the surface of the neutron star. Assuming co-located non-thermal X-ray and $\gamma$-ray emission regions, detection of thermal emission from the polar cap and the lack of non-thermal X-rays indicate that we are viewing the source close to the pole of the neutron star, \ie with small $|\beta|$. Given the absence of detection of non-thermal emission from the pulsar, we cannot determine the parameters of the system like the X-ray efficiency of the pulsar ($\eta_{\rm psr} \equiv L_{\rm psr}/\dot{E}$) or the ratio of the efficiency of the PWN to its pulsar ($\eta_{\rm pwn}/\eta_{\rm psr}$).  Given the spin period and its first derivative ($P$ and $\dot{P}$ ) as determined from radio observations \citep{2003MNRAS.342.1299K} , the equatorial surface magnetic field can be estimated to be  $B \equiv 3.2 \times 10^{19} (P\dot{P})^{\frac{1}{2}} \equiv$ 0.7$\times10^{12}$ G (assuming a dipole magnetic field). This is in the range expected for a neutron star powering a PWN.

\subsection{Morphology and properties of the compact PWN}
The spectrum of the compact nebula is hard, compatible with a power law of $\Gamma=1.12\pm0.25$. This is compatible with that observed in other PWNe \citep{pwnchandra} and is also compatible with predictions of particle wind models assuming relativistic Fermi acceleration \citep{2001MNRAS.328..393A}. 

At a distance of $d=900$~pc, the compact PWN radius $\sim$ 10\arcs\ corresponds to a physical size of 0.06 pc. This is similar to the inner PWN of Vela \citep[$\sim$ 0.1 pc;][]{pavlov2001}. The fainter and diffuse PWN extends much farther as hinted from the \xmm \obs \citep{facero2013} corresponding to a physical size of $\sim$ 0.6 pc. 
An $L_{\rm eff}$  (0.5--8 keV) of $3.3\times10^{31}\ergsec$ indicates a nebular efficiency ($\eta_{\rm pwn} \equiv L_{\rm psr}/\dot{E}$) of $\sim$ 10$^{-5}$, which is similar to what is observed for Vela and some Vela-like pulsars and their PWNe (which have their $\dot{E}$ in the same ballpark range). This is, however, only a lower limit as the contribution from the diffuse nebula has not been taken into account in this case. 
Morphological modelling of the inner PWN with a double-torus model provides a reasonable fit, resulting in an observer angle $\zeta_{\rm PSR} < 37\dg$. The eastern outer tail-like feature which is not accounted for in the spatial model could represent the outer jet of the nebula. In that case the system would have dominance of one of the jet components, which also supports a small value of $\zeta_{\rm PSR}$.

\subsection{Pulsar geometry \& its implications}

Constraints obtained from spatial modelling indicate $\zeta_{\rm PSR}\lesssim 37\dg$ (although with some uncertainty, since we cannot rule out a higher value of $\zeta_{\rm PSR}$ solely from the spatial modelling). Constraints from geometric\footnote{We note that while the geometric models assume constant emissivity in the corotating frame, it is very likely that there will be an azimuthal dependence of this quantity in more detailed models, given such a dependence of the electric field. This should terminate some of the low-level (off-peak) TPC radiation at small $\alpha$ and $\zeta_{\rm PSR}$. More physical models which self-consistently modulates the emissivity via an electric field and associated particle transport and radiation calculations is beyond the scope of the current geometric models.} light curve modelling indicate $\alpha\lesssim55\dg$ and $\zeta\lesssim55\dg$, and $10\dg\lesssim|\beta|\lesssim30\dg$ (due to a single visible radio peak, and from the width of the radio pulse). A $\chi^2$ fit to the radio light curve yields a best fit at $(\alpha,\zeta_{\rm PSR}) = (22\dg, 8\dg)$, with a few alternative fits within $3\sigma$, including one at $(\alpha,\zeta_{\rm PSR}) = (9\dg, 25\dg)$. Assuming (nearly) co-located regions of non-thermal X-ray and $\gamma$-ray emission, this is also consistent with the lack of (pulsed) non-thermal X-rays from the pulsar. To further break the degeneracy of $\alpha$-$\zeta_{\rm PSR}$ parameter space and improve constraints on the individual values of $\alpha$ and $\zeta_{\rm PSR}$, one could pursue radio polarization modelling, as well as X-ray light curve modelling, should radio polarization data become available and X-ray pulsations be detected\footnote{It is also true that should (weak) $\gamma$-ray pulsations be detected in future, this will imply quite large values for $\alpha$ and $\zeta_{\rm PSR}$, while the value of $|\beta|$ should remain small.}. The obtained range of ($\alpha$, $\zeta_{\rm PSR}$) implies we are viewing the system relatively close to the spin axis of the pulsar. Limits on $\beta$ further implies we are also viewing the system relatively close to the magnetic poles of the pulsar.

Interestingly, \citet{R16} recently found that radio profile widths of a sample of 35 young, energetic radio pulsars that are not detected in $\gamma$ rays are typically broader than those of another sample of 35 pulsars which are detected in both radio and $\gamma$ rays, and that the boundary between these two populations increases with $\dot{E}$. The most plausible scenario to qualitatively explain this is the geometry: one needs a small $\alpha$ (to have wide radio pulses) and a small $|\beta|$ (to cross the radio beam and thus detect the radio pulsar), implying a small $\zeta$ (which also, crucially, ensures missing the $\gamma$-ray emission concentrated at the equator, making the pulsar invisible in $\gamma$ rays). This corresponds exactly to our independent arguments used above to constrain the geometry of PSR J0855-4644. \citet{G14} have furthermore argued for small $\zeta_{\rm PSR}$ in the case of nearby, energetic millisecond pulsars that are undetected in $\gamma$ rays, suggesting that the geometric arguments made above hold for the entire pulsar population.

\section{Summary \& Future work}
\label{sec:conc}
We have carried out a detailed spatial and spectral study of the compact PWN and X-ray counterpart of the {\rm fast-spinning} and energetic pulsar \psr. The pulsar is faint and soft in the X-rays (accounts for $\sim$ 1\% of the total emission) surrounded by a brighter, hard and compact PWN of about 0.06 pc in extent. The PWN is elongated along a symmetry axis composed of two lobes more or less equidistant from the pulsar in the middle. The morphology is compatible with either a double-torus morphology although other models (double-torus and jets or jets only) can not be ruled out in the present observation. Spatial modelling of the two lobes of the PWN with a double-torus model captures the gross structure of the nebula and provides an estimate of the observer angle of the pulsar $\zeta_{\rm PSR}$ < $37\dg$. Spectral analysis confirms the hard nature of the PWN spectrum ($\Gamma = 1.12\pm0.25$) with no changes in spectral parameters detected between the outer and inner PWN, or the east and west lobe in the present observation. The pulsar spectrum is compatible with that of a blackbody of  $T_{\rm BB} = 2.0 \times 10^{6}$~K and effective radius $R_{\rm eff}\sim$ 1.5~km, suggesting that the emission originates from the hot spots on the poles of the neutron star.

We find a consistent scenario emerging from constraints on the geometry of the system, derived independently from the X-ray morphological modelling of the nebula and from the radio / $\gamma$-ray light curve modelling. These constraints are compatible with values of $\alpha\lesssim55\dg$ and $\zeta_{\rm PSR}\lesssim55\dg$, and $10\dg\lesssim|\beta|\lesssim30\dg$, (additionally, we find a best-fit radio profile at $(\alpha,\zeta_{\rm PSR}) = (22\dg, 8\dg)$), which implies that we are viewing the pulsar relatively close to one of its magnetic poles, and relatively close to the spin axis of the pulsar. In this scenario, we do not expect to see the $\gamma$-ray pulsations which is supposed to originate from the outer magnetosphere \citep[e.g.,][]{Grenier15} from caustics mostly located near the equatorial regions at large $\zeta_{\rm PSR}$. This proposed scenario (which we do not claim to be the only plausible one) would explain why, despite the high $\dot{E}$/$d^{2}$ value for this system, no $\gamma$-ray pulsations have been detected so far with the \textit{Fermi}-LAT telescope. For co-located regions of non-thermal X-ray and $\gamma$-ray emission, this is also consistent with the lack of non-thermal X-rays from the pulsar.

\psr~is the most energetic pulsar after Vela in the nearby environment (distance $<$ 1 kpc), and therefore has a great appeal in studying the morphology in detail and enhancing our understanding of the PWNe. A future \chandra~\obs will be used to further test for changes in position and brightness of the lobes with respect to the pulsar symmetry axis to disentangle the dominance of either (torus vs.\ jet) components. Obtaining radio polarization data to look for sweeps of the position angle will also be useful to put further constraints on $\beta$. Finally, X-ray pulsations and the light curve shape may lead to independent constraints on the pulsar geometry and mass-to-radius ratio \citep[e.g.,][]{Bogdanov13_J0437}.

%=======================================================================

\begin{acknowledgements}
We gratefully acknowledge fruitful discussions with Alice Harding. This work is based on the research supported wholly / in part by the National Research Foundation of South Africa (Grant Numbers 87613, 90822, 92860, 93278, and 99072). The Grantholder acknowledges that opinions, findings and conclusions or recommendations expressed in any publication generated by the NRF supported research is that of the author(s), and that the NRF accepts no liability whatsoever in this regard.

\end{acknowledgements}

\bibliographystyle{aa} % style aa.bst
\bibliography{j0855b}

\begin{thebibliography}{41}
\expandafter\ifx\csname natexlab\endcsname\relax\def\natexlab#1{#1}\fi

\bibitem[{{Abdo} {et~al.}(2013){Abdo}, {Ajello}, {Allafort}, {Baldini},
  {Ballet}, {Barbiellini}, {Baring}, {Bastieri}, {Belfiore}, {Bellazzini}, \&
  et~al.}]{abdo13-2PC}
{Abdo}, A.~A., {Ajello}, M., {Allafort}, A., {et~al.} 2013, \apjs, 208, 17

\bibitem[{{Acero} {et~al.}(2013){Acero}, {Gallant}, {Ballet}, {Renaud}, \&
  {Terrier}}]{facero2013}
{Acero}, F., {Gallant}, Y., {Ballet}, J., {Renaud}, M., \& {Terrier}, R. 2013,
  \aap, 551, A7

\bibitem[{{Achterberg} {et~al.}(2001){Achterberg}, {Gallant}, {Kirk}, \&
  {Guthmann}}]{2001MNRAS.328..393A}
{Achterberg}, A., {Gallant}, Y.~A., {Kirk}, J.~G., \& {Guthmann}, A.~W. 2001,
  \mnras, 328, 393

\bibitem[{{Arons}(1983)}]{Arons83}
{Arons}, J. 1983, \apj, 266, 215

\bibitem[{{Arzoumanian} {et~al.}(2002){Arzoumanian}, {Chernoff}, \&
  {Cordes}}]{ACC02}
{Arzoumanian}, Z., {Chernoff}, D.~F., \& {Cordes}, J.~M. 2002, \apj, 568, 289

\bibitem[{{Bogdanov}(2013)}]{Bogdanov13_J0437}
{Bogdanov}, S. 2013, \apj, 762, 96

\bibitem[{{Cash}(1979)}]{cash1979}
{Cash}, W. 1979, \apj, 228, 939

\bibitem[{{Cheng} {et~al.}(1986){Cheng}, {Ho}, \& {Ruderman}}]{CHR86a}
{Cheng}, K.~S., {Ho}, C., \& {Ruderman}, M. 1986, \apj, 300, 500

\bibitem[{{Cheng} {et~al.}(2000){Cheng}, {Ruderman}, \& {Zhang}}]{Cheng00}
{Cheng}, K.~S., {Ruderman}, M., \& {Zhang}, L. 2000, \apj, 537, 964

\bibitem[{{Daugherty} \& {Harding}(1996)}]{1996ApJ...458..278D}
{Daugherty}, J.~K. \& {Harding}, A.~K. 1996, \apj, 458, 278

\bibitem[{{Deutsch}(1955)}]{Deutsch55}
{Deutsch}, A.~J. 1955, Annales d'Astrophysique, 18, 1

\bibitem[{{Dyks} {et~al.}(2004){Dyks}, {Harding}, \& {Rudak}}]{Dyks04}
{Dyks}, J., {Harding}, A.~K., \& {Rudak}, B. 2004, \apj, 606, 1125

\bibitem[{{Dyks} \& {Rudak}(2003)}]{Dyks03}
{Dyks}, J. \& {Rudak}, B. 2003, \apj, 598, 1201

\bibitem[{{Gaensler} \& {Slane}(2006)}]{gaensler2006}
{Gaensler}, B.~M. \& {Slane}, P.~O. 2006, \araa, 44, 17

\bibitem[{{G{\"a}nsicke} {et~al.}(2002){G{\"a}nsicke}, {Braje}, \&
  {Romani}}]{2002yCat..33861001G}
{G{\"a}nsicke}, B.~T., {Braje}, T.~M., \& {Romani}, R.~W. 2002, VizieR Online
  Data Catalog, 338

\bibitem[{{Goldreich} \& {Julian}(1969)}]{GJ69}
{Goldreich}, P. \& {Julian}, W.~H. 1969, \apj, 157, 869

\bibitem[{{Gonthier} {et~al.}(2004){Gonthier}, {Van Guilder}, \&
  {Harding}}]{Gonthier04}
{Gonthier}, P.~L., {Van Guilder}, R., \& {Harding}, A.~K. 2004, \apj, 604, 775

\bibitem[{{Grenier} \& {Harding}(2015)}]{Grenier15}
{Grenier}, I.~A. \& {Harding}, A.~K. 2015, Comptes Rendus Physique, 16, 641

\bibitem[{{Guillemot} \& {Tauris}(2014)}]{G14}
{Guillemot}, L. \& {Tauris}, T.~M. 2014, \mnras, 439, 2033

\bibitem[{{Harding} {et~al.}(2007){Harding}, {Grenier}, \&
  {Gonthier}}]{Harding07_Geminga}
{Harding}, A.~K., {Grenier}, I.~A., \& {Gonthier}, P.~L. 2007, \apss, 309, 221

\bibitem[{{Helfand} {et~al.}(2001){Helfand}, {Gotthelf}, \&
  {Halpern}}]{helfand2001}
{Helfand}, D.~J., {Gotthelf}, E.~V., \& {Halpern}, J.~P. 2001, \apj, 556, 380

\bibitem[{{Hessels} {et~al.}(2004){Hessels}, {Roberts}, {Ransom}, {Kaspi},
  {Romani}, {Ng}, {Freire}, \& {Gaensler}}]{hessels2004}
{Hessels}, J.~W.~T., {Roberts}, M.~S.~E., {Ransom}, S.~M., {et~al.} 2004, \apj,
  612, 389

\bibitem[{{Johnson} {et~al.}(2014){Johnson}, {Venter}, {Harding}, {Guillemot},
  {Smith}, {Kramer}, {{\c C}elik}, {den Hartog}, {Ferrara}, {Hou}, {Lande}, \&
  {Ray}}]{Johnson2014}
{Johnson}, T.~J., {Venter}, C., {Harding}, A.~K., {et~al.} 2014, \apjs, 213, 6

\bibitem[{{Kargaltsev} \& {Pavlov}(2008)}]{pwnchandra}
{Kargaltsev}, O. \& {Pavlov}, G.~G. 2008, in American Institute of Physics
  Conference Series, Vol. 983, 40 Years of Pulsars: Millisecond Pulsars,
  Magnetars and More, ed. C.~{Bassa}, Z.~{Wang}, A.~{Cumming}, \& V.~M.
  {Kaspi}, 171--185

\bibitem[{{Kijak} \& {Gil}(2003)}]{Kijak03}
{Kijak}, J. \& {Gil}, J. 2003, \aap, 397, 969

\bibitem[{{Kramer} {et~al.}(2003){Kramer}, {Bell}, {Manchester}, {Lyne},
  {Camilo}, {Stairs}, {D'Amico}, {Kaspi}, {Hobbs}, {Morris}, {Crawford},
  {Possenti}, {Joshi}, {McLaughlin}, {Lorimer}, \&
  {Faulkner}}]{2003MNRAS.342.1299K}
{Kramer}, M., {Bell}, J.~F., {Manchester}, R.~N., {et~al.} 2003, \mnras, 342,
  1299

\bibitem[{{Lampton} {et~al.}(1976){Lampton}, {Margon}, \& {Bowyer}}]{Lampton76}
{Lampton}, M., {Margon}, B., \& {Bowyer}, S. 1976, \apj, 208, 177

\bibitem[{{Muslimov} \& {Harding}(2003)}]{Muslimov2003}
{Muslimov}, A.~G. \& {Harding}, A.~K. 2003, \apj, 588, 430

\bibitem[{{Ng} \& {Romani}(2004)}]{ngromani2004}
{Ng}, C.-Y. \& {Romani}, R.~W. 2004, \apj, 601, 479

\bibitem[{{Ng} \& {Romani}(2008)}]{ngromani2008}
{Ng}, C.-Y. \& {Romani}, R.~W. 2008, \apj, 673, 411

\bibitem[{Parzen(1962)}]{Parzen62}
Parzen, E. 1962, Ann. Math. Statist., 33, 1065

\bibitem[{{Pavlov} {et~al.}(2001){Pavlov}, {Kargaltsev}, {Sanwal}, \&
  {Garmire}}]{pavlov2001}
{Pavlov}, G.~G., {Kargaltsev}, O.~Y., {Sanwal}, D., \& {Garmire}, G.~P. 2001,
  \apjl, 554, L189

\bibitem[{{Pierbattista} {et~al.}(2015){Pierbattista}, {Harding}, {Grenier},
  {Johnson}, {Caraveo}, {Kerr}, \& {Gonthier}}]{Pierbattista15}
{Pierbattista}, M., {Harding}, A.~K., {Grenier}, I.~A., {et~al.} 2015, \aap,
  575, A3

\bibitem[{{Rankin}(1993)}]{Rankin93}
{Rankin}, J.~M. 1993, \apj, 405, 285

\bibitem[{{Romani}(1996)}]{Romani96}
{Romani}, R.~W. 1996, \apj, 470, 469

\bibitem[{{Romani} \& {Yadigaroglu}(1995)}]{1995ApJ...438..314R}
{Romani}, R.~W. \& {Yadigaroglu}, I.-A. 1995, \apj, 438, 314

\bibitem[{{Rookyard} {et~al.}(2016){Rookyard}, {Weltevrede}, {Johnston}, \&
  {Kerr}}]{R16}
{Rookyard}, S.~C., {Weltevrede}, P., {Johnston}, S., \& {Kerr}, M. 2016, ArXiv
  e-prints

\bibitem[{Rosenblatt(1956)}]{Rosenblatt56}
Rosenblatt, M. 1956, Ann. Math. Statist., 27, 832

\bibitem[{{Venter} {et~al.}(2009){Venter}, {Harding}, \&
  {Guillemot}}]{Venter09}
{Venter}, C., {Harding}, A.~K., \& {Guillemot}, L. 2009, \apj, 707, 800

\bibitem[{{Weisskopf} {et~al.}(2000){Weisskopf}, {Hester}, {Tennant}, {Elsner},
  {Schulz}, {Marshall}, {Karovska}, {Nichols}, {Swartz}, {Kolodziejczak}, \&
  {O'Dell}}]{weisskopf2000}
{Weisskopf}, M.~C., {Hester}, J.~J., {Tennant}, A.~F., {et~al.} 2000, \apjl,
  536, L81

\bibitem[{{Zavlin}(2009)}]{zalvin2009}
{Zavlin}, V.~E. 2009, in Astrophysics and Space Science Library, Vol. 357,
  Astrophysics and Space Science Library, ed. W.~{Becker}, 181

\end{thebibliography}

\end{document}